%Paper: cond-mat/9503034
%From: "ROBERT HOLYST" <HOLYST@ichf.edu.pl>
%Date: Tue, 7 Mar 1995 14:42:20 MEZ-1

\magnification=1200
\vsize=8.5truein
\hsize=6truein
\baselineskip=20pt
\centerline{\bf Dislocations in uniaxial lamellar phases of liquid crystals,}
\centerline{\bf polymers and amphiphilic systems}
\vskip 20pt
\centerline{by}
\vskip 20pt
\centerline{Robert Ho{\l}yst$^1$ and Patrick Oswald$^2$}
\vskip 20pt
\centerline {$^1$Institute of Physical Chemistry of the
Polish Academy of Sciences, Dept. III,}
\centerline {Kasprzaka 44/52,
01224 Warsaw, Poland}
\centerline{$^2$ Ecole Normale Superieure de Lyon, Laboratoire de Physique}
\centerline{46, All\'ee d'Italie-f-69364 Lyon Cedex 07, France}
\vskip 20pt
\centerline{\bf{Abstract}}
Dislocations in soft condensed matter systems such as lamellar systems of
polymers, liquid crystals
and ternary mixtures of oil, water and surfactant (amphiphilic systems)
are described in the
framework of continuum elastic theory. These systems are
the subject of studies of physics, chemistry and biology. They also find
applications in the industry. Here we will discuss in detail the
influence of dislocations on the bulk and surface properties of these lamellar
phases. Especially the latter properties have only been recently
studied in detail.
We will present the experimental evidence of the existence of
screw and edge
dislocations in the systems and study their static properties such
as: energy, line
tension and core structure. Next we will show how does the surface
influence the
equilibrium position of dislocations in the system. We will give the
theoretical predictions and present the experimental results on thin
copolymer films, free standing films of liquid crystals and smectic droplets
shapes. In semi-infinite
lamellar systems characterized by small surface tension the
dislocation is stabilized at a finite distance, $h_{eq}$, from the surface,
due to the surface bending elastic constant,$K_s$ (for zero surface tension
$h_{eq}\approx K_s/2 K$, where $K$ is the bulk bending elastic constant).
For large surface tension the edge dislocations are strongly repelled
by the surface and the equilibrium location for finite symmetric systems
such as
free standing liquid crystal films shifts towards the center of the system.
The surface is deformed by dislocations. These deformations are known
as edge profiles. They will be discussed for
finite systems with small and large surface tension.
Surface deformations induce elastic interactions between
edge dislocations,
which decay exponentially with distance with decay length proportional to
$\sqrt{D}$, where $D$ is the size of the system normal to lamellas.
Two screw dislocations in finite system interact with the logarithmic
potential, which is proportional to the surface tension and inversly
proportional to $D$.
The surface induced elastic interactions will be compared to, well known,
bulk deformation induced
interactions.
A new phenonenon discussed in our paper is the
fluctuations induced interactions between
edge dislocations, which follows from the Helfrich mechanism
for flexible objects. At suitable conditions edge dislocations can undergo
an unbinding transition. Also a single dislocation loop
can undergo an unbinding transition. We will calculate the
properties of the loop inside finite system and discuss in
particular the unbinding transition in freely suspended
smectic films. We shall also compute the equilibrium
size of the loop contained between two hard walls.
Finally we will discuss
the dynamical bulk properties of dislocations such as: mobility (climb
and glide),
permeation, and helical instability of screw dislocations.
Lubrication theory will also be discussed.

PACS numbers: 61.30.Jf, 61.72.Lk, 61.41.+e, 61.30.Cz
\vfill\eject
\centerline{\bf I. Introduction}

Before we enter the main subject of this review, which is the edge and screw
dislocations
in soft condensed matter systems such as smectic liquid crystals, lamellar
phases of diblock copolymers
and lamellar phases in ternary mixtures of oil, water and surfactants
(amphiphilic system)
we would like
to put the subject of dislocations in a broader perspective.
The phenomena of
upheavals of the mountains, slipping of glaciers
from the high mountains and easy deformations of metals
are all controlled by the defects known as dislocations.
In the defects free samples of solids the shear displacement requires
(as theoretically calculated)
shears 5 order of magnitude larger than is routinely applied in typical
deformation processes  (for an Introduction to
dislocations in solids see$^{1,2}$). This fact aroused the curiosity of
scientists and in 1936
G.I.Taylor, E.Orowan and M.Polanyi resolved the problem
of deformations by postulating that deformations of solids are controlled by
crystal imperfections (defects) called dislocations.  It took almost 20 years
before scientists observed dislocations in solid. One  of the first
observations
were done in 1953 by Hedges and Mitchell and later in 1956 by Hirsch, Horne
and Whelan and independently by Bollmann. The two techniques used in the
experiments were: the decoration technique with optical observation
and the transmission electron microscopy. In the decoration technique the
guest atoms in a solid locate themselves along the
dislocation line providing contrast for optical observations$^{3}$.
In the transmission
electron microscopy (TEM) the diffraction and interference of electrons$^4$
allows to see
the dislocations. Strong scattering of electrons limits the
application of TEM
to rather thin samples.

Dislocations also appear in liquid
partially ordered lamellar phases(Figs.1-3). The molecules in lamellar phases
are arranged in liquid like layers parallel to each other. Although the
symmetry of these phases is the same in amphiphilic systems$^5$,
diblock copolymers$^6$ and liquid crystals$^7$, their structure parameters
can be very different. In liquid crystals the period of the lamellar
structure, d,
is set by the size of the molecule which is roughly 30\AA;
in amphiphilic system the amount
of surfactant, water and/or oil allows to vary the lamellar size
from $\sim$20\AA (for large concentration of surfactant) to $\sim$1000\AA (for
very small concentration of surfactant)$^8$.
In block copolymers the period is determined by
the size of the polymer molecule.
In the strong segregation limit when the polymer
molecules are strongly stretched
in the lamellas the period is given by $M^{2/3}l$, where M is the number of
statistical segments of a polymer and $l$ is the length of a segment,
thus the period can be as large as few hundred \AA.
Also the elastic constants characterizing the lamellar phases can be very
different for
these systems and as we will show the behavior of dislocations is sensitive
to the value of
elastic parameters characterizing a system. In the following
we shall restrict our discussion to uniaxial lamellar phases, i.e.
smectic A liquid crystals and
$L_\alpha$ phases of lyotropics.
Two types of dislocations, namely screw and edge dislocations,
are shown schematically in Fig.4.

The bulk properties of dislocations have been very thoroughly
studied$^{9-11}$
since 1972, when de Gennes presented the equation for elastic deformations
induced by a single
edge dislocation in the infinite lamellar system$^{7,12}$. The first
observation of
elementary edge dislocation has been done by Meyer et al in 1978$^{13}$,
although in earlier experiments the dislocations
of large Burgers vector have already been seen$^{14-18}$.
The influence of surfaces on dislocations has been first studied by
Pershan$^{19}$.
He found that dislocations are repelled from solid surfaces and attracted
by a free surface as in solids. The latter conclusion is erroneous as
shown by recent calculations$^{20}$ and experiments$^{21,22}$.
Indeed, surface tension effects cannot be neglected at the air - smectic
interface, contrary to Pershan's assumption.
For instance, recent observations by TEM (transmission
electron microscopy)$^{22}$ and AFM (atomic force microscopy)$^{21,22}$
of ultra-thin films of A-B diblock copolymers deposited on a solid
substrate showed that elementary edge dislocations stabilize within the film
and are
repelled from the free surface. Moreover Pershan analysis failed to
explain the shape of the surface deformations induced by dislocations.
The repulsion of the dislocation from the surface is also observed  in
free standing
films$^{23}$ and in the ``gouttes \`a gradins'' of Grandjean$^{24}$.
The repulsion from the free surface of the giant dislocations
separating two terraces of a ``goutte \`a gradins'' was first pointed
out by C.Williams$^{25}$ in his thesis in 1976.  Williams
also emphasized that surface tension was responsible for this repulsion.

One of the aims of this review is thus the discussion of the recent progress
in the dislocation theory for systems bounded by surfaces.
The list of problems includes: surface deformations (edge profiles),
equilibrium location of dislocations near surfaces and surface induced
interactions between edge and screw dislocations.
We shall also discuss some of their bulk, static  properties such as
the core structure and the formation of the giant dislocations
(which do not exist in solids) and some consequences of their flexibility
on their interactions. Finally we will analyse the
dynamic
properties of dislocations. These analysis
will include the mobility, their instability under mechanical
stress and their role in microplasticity and rheology.

The plan of the paper is as follows:  In section II
we present the experimental evidences for
the existence of dislocations in lamellar phases.
In section III we discuss the
elastic theory for lamellar phases and in section IV we briefly
recall the hydrodynamic equations governing their
dynamical properties. In section V the distortions
induced in the bulk by edge and screw dislocations are calculated.
Here we also calculate the energy and line tension of dislocations.
In section VI we find the equilibrium location of dislocation
in finite system bounded by surfaces. Here we discuss both the theory
and experiments. The edge profiles are calculated in section VII
and elastic interactions
 are calculated in section VIII. Sections IX, X and XI are devoted to the
flexibility of dislocation line and unbinding transition
in the system of dislocations.
Finally in sections XII, XIII and XIV the
dynamical properties of dislocations are discussed.
Conclusions are contained in section XV.

\centerline{\bf II. Experimental observations of dislocations}

One of the main problem in optical detection of defects
is the sufficient optical contrast against the lamellar background.
In 1978 Meyer et al.$^{13}$ observed an array of elementary
edge dislocations in
a wedge shaped sample of smectic liquid crystal, contained between
two glass plates. In order to increase the optical contrast
they used the phenomenon of
phase transition between smectic-A, where molecules are roughly perpendicular
to layers,
and smectic-C, where liquid crystalline molecules are tilted with respect to
layers. Tilting is accompanied by the decrease of the
layer thickness and thus can relieve stress
in a wedge shaped sample. Because of the wedged shape the
thickness of the sample changes and the stress induces
formation of dislocations in a sample as shown
schematically in Fig.5. For temperatures close to the
smectic-A smectic-C transition temperature the regions close to the
dislocations are tilted. Thus under the polarizing microscope one can get
dark regions separated by bright regions, depending on the tilt and the
angle between the  crossed polarizers.
The distance between the dislocations in the array, $l$ is given by
$l=d/\alpha$, where $\alpha$ is a small angle between the glass plates.
The accurate measurements of the angle were done by the Fabry-Perot
interferometer. In this method one cannot tell wheather dislocations
are located at the surface or in the bulk, but the fact that only
dislocations satisfying the geometrical constraint given by $l$
remain after annealing  suggests that they are free to move and are not
attached to the solid surface. This is confirmed by creep experiments
discussed in section XIII.

Another optical method of observation based on fluorescence
has been applied to
phospholipid lamellar phase (example of self assembling amphiphilic system)
with alternate layers of water and fluid-like bilayers of phospholipid$^{26}$.
The system is placed in the same wedge shaped
container and the phospholipid is doped with fluorescent, lipid
analogue molecule.
The sample is annealed for some time (several weeks) and next
excited with a laser.
The fluorescent intensity is monitored as a function of position as shown in
Fig.6. The visible steps are clear signs of parallel, elementary edge
dislocations.
The long annealing time
that is necessary to reach equilibrium in this system
is related to permeation (flow of molecules across lamellas,
Sections XII, XIII). Permeation is much more difficult in
lyotropic systems like this one than in thermotropic systems studied by
Meyer et al$^{13}$, where the annealing times were much shorter.

The exact location of dislocation in a sample can be determined by TEM$^{22}$.
Here we briefly describe this experiment$^{21,22}$ as shown schematically
in Fig.7. The ultra-thin films (thickness $D\sim$ 1000\AA )
of lamellar phase of AB diblock copolymers (period $d\approx 315$\AA)
are deposited on solid substrate.
The films are first prepared in the disordered solid state and heated
above their glass transition but below the order-disorder transition.
The lamellar (smectic) ordering appears with layers parallel to the
substrate. The free surface of the film is not flat since circular steps
of micrometer size are formed, separated by equally large holes.
The process of the formation of the holes and steps
is associated with the nucleation of dislocations inside the film and
is due to the incompatibility of the initial thickness of the
disordered film, $D$, with the smectic period of the final lamellar phase
i.e. $Nd<D<(N+1) d$. Because $N+1$ is the number of layers below the step
and $N$ is the number of layers below the hole the height of the step is
equal to the lamellar period, $d$.
Indeed applying TEM one can observe the dislocation itself, while by applying
AFM one can measure the shape (size, height)
of the step and the width of the profile at the edge of steps (edge profiles).
The slices (cross section shown in Fig.7)
of thickness of 500{\AA} are prepared with layers
parallel to the electron beam$^{22}$.
To have a contrast between the PS and PBMA domains PS are stained
by treating it with RuO$_4$ vapor.
The electrons are diffracted on the
layers and the micrographs shows
the positions of layers and in particular of defects.
This is a direct way of seeing dislocations.
The slices have to be thin, because of strong interactions between
electrons and matter. The originally liquid
samples are of course vitrified before observation. Since the polymers are
very viscous fluids, the structure remains intact in the
process of vitrification.

Edge dislocations are also visible in free standing films$^{23}$.
They are usually obtained by smearing a smectic phase with a wiper across
a fixed aperture made in a solid substrate. This technique used by most
researchers and invented by Friedel 70 years ago$^{27}$ has been considerably
improved by P.Piera\'nski$^{23}$.
He used  a rectangular frame of {\bf variable}
area, allowing precise control (up to one layer) of the film thickness
in the process of preparation. Fig.8a shows a schematic structure of the
film near the edge of the aperture. As we see the dislocations are located
in the middle of the smectic meniscus.
Fig.8b shows a typical arch - texture obtained when the film is quickly
stretched. Here the film is divided into areas of different thickness
which are separated by lines of circular shape. In general the thickness
variation between two adjacent regions is one layer. Some lines are
linked by knots too. Piera\'nski $^{23}$ showed that these  lines are
bulk dislocations located in the middle of the film and suggested that there
exists some dust particles (not always visible under the
microscope) at each junction
(the knots) between the dislocations (Fig.8c).

Edge dislocations also play a crucial role in the problem of the equilibrium
shape of a smectic A droplet
deposited on a glass plate with strong homeotropic
anchoring (molecules arranged perpendicularly to the substrate)$^{28,29}$.
Droplets shapes can be measured very precisely ($\pm$50\AA) by
Michelson interferometry. This experiment  was performed with materials that
have either a nematic-smectic A phase transition (8OCB, 4O.8) or
an isotropic-smectic A phase transition. It revealed that small droplets
(usually less than 200 micrometers in diameter) have a single facet parallel
to the substrate, while larger droplets are shaped like a spherical cap.
Fig.9  shows typical profiles of a small 8OCB droplet, when the temperature
is decreased below $T_{NA}$ (nematic-smectic phase transtion temperature).
Because this transition is second order, the size
of the facet vanishes at $T_{NA}$
and increases like
$(T_{NA}-T)^{0.45}$ when the temperature is decreased below
$T_{NA}$. In 10OCB where the first order
smectic-A isotropic phase transition occurs, the facet has a finite size
at the transition. If the radius of the droplet is larger than 200
micrometers, the top facet disappears and the droplet becomes spherical
as in the previous case. It means that  steps on the free surface
associated with facets (Fig.10a) are replaced by the dislocations
distributed throughout the interior of the droplet (Fig.10b).
This is possible since dislocations are repelled by both, the solid
substrate and the free surface. In very large droplets,  dislocations
can group together and form giant dislocations separated by  terraces
parallel to the glass plate (``gouttes \`a gradins'' Fig.11)

The shapes of the droplets has been explained by the theory, where
interactions between the steps and between the dislocations, together with
the surface and finite size  effects
have been properly taken into account$^{30}$.
These finite size efects are also
important to explain the appearances in some materials
(4O.8) of a secondary facet around the primary one (Fig.12). This facet
occurs at low temperatures and in large enough droplets (diameter more than
100 micrometers). This means that very special conditions must be fulfilled
simultaneously  in order to observe it.
This unexpected facet (there is no  terrace periodicity along
the new facet direction) results from the elastic attraction between like
steps, balanced by entropic repulsion. We shall see in section VIII that
there can be an attraction between
like  dislocations  (of the same sign)
in films, similar to the attraction between
steps as observed in this
experiment.

The screw dislocations have been also observed in
self assembled amphiphilic systems in TEM$^{16,31,32}$ experiments
(Fig. 13).
The density of screw dislocations can be as large as $10^8$
per cm$^2$ in many lyotropic systems. High density suggests that their energy
is rather small, a fact also established in the theory (Section V).

\centerline{\bf III. Elastic theory of lamellar phases}

The lamellar phase consists of parallel, equidistant
two dimensional liquid layers. The average distance between the layers
is denoted $d$.
For convenience we assume
that unperturbed layers are perpendicular to the z-axis.
The layers deform easily and their deformations are conveniently
described by
two unit vectors, one normal to the layer,
$\hat {\bf m}({\bf r})$ and one describing the average orientation of the
molecules $\hat{\bf n}({\bf r})$
(called in liquid crystals the director), and a scalar quantity, which
measures the distance between the layers along the normal.
We further assume that other quantities like the nematic order
parameter and the density (for liquid crystals) or surfactant and water
concentrations (for amphiphilic system) adjust to the layer deformations.
In the first approximation, density changes do not generate new
terms in the distortion energy, but only renormalize the lamellar
elastic constants$^{7,33}$. Using the two aforementioned
quantities we can write the distortion energy density
in the simplest form invariant
with respect to global rotations:
$$\eqalign{f_b={1\over{2}}
&B\left({{d-d_0}\over {d_0}}\right)^2+
{1\over 2}K_1\vert\nabla \hat{\bf n}({\bf
r})\vert^2+{1\over 2}K_2\vert\hat{\bf n}({\bf
r})\cdot\nabla\times \hat{\bf n}({\bf
r})\vert^2\cr+{1\over 2}&K_3\vert\hat{\bf n}({\bf
r})\times (\nabla\times \hat{\bf n}({\bf
r}))\vert^2+{1\over 2}B_1(\hat{\bf n}-\hat{\bf m})^2,\cr}\eqno(3.1)$$
where $d_0$ is the unperturbed layer spacing, B is the
Young modulus of the layers
associated with the variations of the thickness and $B_1\sim B$
is the elastic modulus associated with the deviations of
the average orientation
of molecules from the normal to the layers$^{34}$.
The divergence and the
curl operators in Eq.(3.1)
are taken with respect to variables in the distorted state
and the point {\bf r} is uniquely related to some other point
${\bf r}_0$ on the unperturbed layer.
We neglect the higher order
terms in $d-d_0$ and in the derivatives of $\hat{\bf n}$.
The last three terms in
this distortion energy have their analogs in nematic liquid crystals.
$K_1$, $K_2$ and $K_3$ are the elastic constants for splay, twist and bend
deformations of $\hat{\bf n}$, respectively.
As one can see the splay of $\hat{\bf n}$
corresponds to the bend of the layers
while the bend of $\hat{\bf n}$ corresponds
to the splay of the layers. From now on we shall use the
notion bend and splay
in connection to layers.
For the last term in Eq(3.1) one finds
that $(\hat{\bf n}-\hat{\bf m})^2\sim \lambda/R$
where $\lambda=\sqrt{K_1/B}$ is roughly proportional for most of the
lamellar systems to $d_0$ and $R$ is a typical radius of curvature
for the director field. Consequently for $R\gg\lambda$ we can set
$\hat{\bf n}=\hat{\bf m}$. We shall keep  in mind however that this
assumption is not valid close to the dislocation core.
Nonetheless from now on we neglect the last term in Eq(3.1).

Since at this point we are interested in the local distortion
energy density in the lamellar phase we can
neglect dislocations, which means that the total number of layers
crossed along any path going from
some point A to another point B is constant i.e. for a closed path
$$\oint{{\hat{\bf n}}\over d}d{\bf l}=0,\eqno(3.2)$$
This is equivalent to the condition$^7$:
$$\nabla\times {{\hat{\bf n}}\over d}=0,\eqno(3.3)$$
where $d$, in general, depends on the position
${\bf r}$. Condition(3.3) eliminates the twist
term from the distortion energy,
but not the splay term ($K_3$). Here we only note that splay of  layers
does not affect
the long wavelength properties of the system.
Combining Eq(3.1) and Eq(3.3) we get
$$f_b={1\over{2}}\left(
B\left({{d-d_0}\over {d_0}}\right)^2+
K_1\left\vert\nabla \hat{\bf n}({\bf
r})\right\vert^2+K_3\left(\left\vert{1\over d}\nabla{d}\right\vert^2
-\left\vert{{\hat{\bf n}({\bf
r})}\over d}\nabla {d}\right\vert^2\right)\right)\eqno(3.4)$$
It is important to notice that a dislocation should not affect the form
of the
distortion energy density, which is a local quantity, whereas dislocation is
described by the global condition i.e. nonvanishing of (3.2).
One can also verify this by direct
calculations$^{7}$. For the edge dislocation along $y$  direction
$\hat{\bf n}$ has only $x,z$ components
and does not depend on $y$ and $\nabla\times \hat{\bf n}/d$
is along the $y$ direction. Thus the twist term drops out in this case.

In lamellar phases,
$\hat{\bf n}$ and $d$ are not independent and can be expressed
in terms of the vertical displacement field $u({\bf r}_0)$.
Although $u$ is a multivalued function in the presence of dislocation,
we can still use it away from the dislocation.
The vector normal to the layer at point ${\bf r}={\bf r}({\bf r}_0)$
is simply given by:
$$\hat{\bf n}={{\left(-{{\partial u}\over{\partial x_0}},
-{{\partial u}\over {\partial y_0}}, 1\right)}\over {\sqrt{1+
\vert\nabla_\bot^{(0)}u\vert^2}}},\eqno(3.5)$$
where ${\bf r}_0=(x_0, y_0, z_0)$ is the coordination point in the unperturbed
system.
The distance between layers measured along $\hat{\bf n}$ is
$$d=d_z(\hat{\bf n}\hat{\bf z}),\eqno(3.6)$$
where $d_z$ is the distance between layers measured along the z-axis.
We find, to the lowest order in the derivatives of u, that
$$d_z=d_0\left(1+{{\partial u}\over {\partial z_0}}\right).\eqno(3.7)$$
which together with Eqs(3.5-6) gives $d$ in terms of $u$,
$$d=d_0{{\left(1+{{\partial u}\over {\partial z_0}}\right)}\over
{\sqrt{1+
\vert\nabla_\bot^{(0)}u\vert^2}}}.\eqno(3.8)$$
In order to represent the distortion energy given by Eq(3.4) in terms
of $u$, we have to transform the point ${\bf r}$ on a perturbed layer
to the point ${\bf r}_0$ on the unperturbed layer and the corresponding
nabla operator as follows:
$$x=x_0,$$
$$y=y_0,$$
$$z=z_0+u(x_0,y_0,z_0),$$
and
$${{\partial}\over {\partial x}}={\partial\over{\partial x_0}}-
{{{\partial u}\over{\partial x_0}}
\over{\left(1+{{\partial u}\over {\partial z_0}}\right)}}{\partial\over
{\partial z_0}},\eqno(3.9)$$
$${{\partial}\over {\partial y}}={\partial\over{\partial y_0}}-
{{{\partial u}\over{\partial y_0}}
\over{\left(1+{{\partial u}\over {\partial z_0}}\right)}}{\partial\over
{\partial z_0}},\eqno(3.10)$$
$${{\partial}\over {\partial z}}=
{{{1}}
\over{\left(1+{{\partial u}\over {\partial z_0}}\right)}}{\partial\over
{\partial z_0}}.\eqno(3.11)$$
Combining these equations we recover the condition given by Eq(3.3)
for the constant number of layers. Finally we obtain the
following expression for the distortion energy density$^{35}$:
$$f_b={1\over{2}}\left(
B\left({{d-d_0}\over {d_0}}\right)^2+
K_1\left\vert\nabla_\bot^{(0)} \hat{\bf n}_\bot
\right\vert^2+K_3\left(\left\vert{1\over d}
\nabla_\bot^{(0)}{d}\right\vert^2
-\left\vert{{\hat{\bf n}_\bot }\over d}\nabla_\bot^{(0)} {d}\right\vert^2
\right)\right),
\eqno(3.12)$$
where $\hat{\bf n}_\bot=-\nabla_\bot^{(0)}u/
\sqrt{1+\vert\nabla_\bot^{(0)} u\vert^2}$ and
$\nabla_\bot^{(0)}=(\partial /\partial x_0, \partial /\partial y_0)$
is the two dimensional nabla operator.
As noted before, although the splay term associated with $K_3$ gives a
coupling between
compression and undulation of layers, it does not affect the
long-wavelength properties
of the system. Also it can be neglected in comparison to the
compression term. Ignoring the splay term and anharmonic
contribution to the bend term, we
find an approximate form of $f_b$:
$$f_b={1\over{2}}\left(B\left({{1+{{\partial u({\bf r_0})}\over
{\partial z_0}}}\over {\sqrt{1+\vert\nabla_\bot^{(0)} u({\bf r}_0)
\vert^2}}}-1\right)^2+
K_1(\triangle^{(0)}_\bot
u({\bf r_0}))^2\right).\eqno(3.13)$$
Here $\triangle^{(0)}_\bot$ is the two-dimensional Laplacian with
respect to $x_0, y_0$.
The distortion energy density given by Eq(3.13)
is invariant with respect to rotations. For example, by
rotating the system
around y-axis by an angle $\theta$, we find:
$$u=z_0\left({1\over {\cos{\theta}}}-1\right)-x_0\tan{\theta}.\eqno(3.14)$$
For such spurious deformation we find from Eq(3.13)
that $f_b=0$ as it should.
Expanding Eq(3.13) in $\vert\nabla_\bot^{(0)}u\vert^2$
we recover the de Gennes distortion energy density$^{7,36}$
$$f_b={1\over{2}}
\left(B\left({{\partial u({\bf r_0})}\over {\partial z_0}}-
{1\over 2}\vert\nabla_\bot^{(0)} u({\bf
r_{0}})\vert^2\right)^2+K_1(\triangle_\bot^{(0)}
u({\bf r_{0}}))^2\right),\eqno(3.15)$$
in the lowest
order of the expansion, losing however the rotational
invariance of $f_b$.
Alternatively, the distortion energy density can be expressed in terms of
$\nabla u$. We find
$$f_b={1\over{2}}\left(B\left({1\over{\sqrt{1-2\left(\partial u/\partial z
-\vert\nabla u({\bf r})\vert^2/2\right)}}}-1\right)^2+
K_1\vert\nabla{\bf\hat n}({\bf r})\vert^2\right)\eqno(3.16)$$
where
$${\bf\hat n}({\bf r})={(-\partial u/\partial x,-\partial u/\partial y,
1-\partial u/\partial z)\over\sqrt{1-2\left(\partial u/\partial z
-\vert\nabla u({\bf r})\vert^2/2\right)}}\eqno(3.17)$$
and we have neglected the splay term.
One easily checks that $f_b=0$ when we rotate the system, i.e.
$$u=z(1-\cos\theta)-x\sin\theta.\eqno(3.18)$$
The transformation properties of $u$
result from the fact that it is defined as
the {\it vertical displacement} of the layer from its
rest position.
 In the lowest order expansion in
$\partial u/\partial z-\vert\nabla u\vert^2/2$ we recover the Grinstein
and Pelcovits expression$^{37}$:
$$f_b={1\over{2}}
\left(B\left({{\partial u({\bf r})}\over {\partial z}}-
{1\over 2}\vert\nabla u({\bf
r})\vert^2\right)^2+K_1(\triangle_\bot
u({\bf r}))^2\right),\eqno(3.19)$$
If we neglect the anharmonic terms we find the harmonic approximation
of the distortion energy:
$$f_b={1\over{2}}\left(B\left({{\partial u({\bf r})}\over
{\partial z}}\right)^2+
K(\triangle_\bot
u({\bf r}))^2\right).\eqno(3.20)$$
Here we define K=K$_1$.
Once again  we note that Eq(3.20)  is valid far away from the dislocation
core.
This form of the bulk distortion
energy will be used in our description of deformations induced by
dislocations in a lamellar (smectic) system.
At the harmonic level of approximation
the form of the distortion energy is the same in
coordinates of both, the deformed and undeformed state
(Eq(3.15), Eq.(3.19)).
The role of the anharmonic terms in lamellar phases
for long wavelength distortions has been
discussed in detail in Refs(7,37). In general its role for
distortions induced by dislocations is not known.

\centerline{\bf IV. Hydrodynamics of smectics: elastic and viscous stress
tensor}

The motion of dislocations can be described in the framework of
the hydrodynamical equations for the lamellar phases. Here we recall
the basic ingredients of the theory$^{7,33}$. Let {\bf v} be the
average velocity of the molecules. If the velocity of
the dislocations is much smaller than the velocity of the first sound,
then the medium can be regarded as incompressible i.e.
$$\nabla{\bf v}=0\eqno(4.1)$$
The momentum conservation equation for the system is
as follows.
$$\rho{D{\bf v}\over{D t}}=\nabla{\bf\sigma}+{\bf F},\eqno(4.2)$$
where $\rho$ is the density, ${\bf\sigma}$ is the stress  tensor
and ${\bf F}$ the bulk forces (gravitation for example). The stress tensor
can be decomposed into three terms:
$${\bf\sigma}=-P{\bf I} + {\bf\sigma^E}+{\bf\sigma^V}.\eqno(4.3)$$
Here P is the hydrostatic pressure (given by condition (4.1)),
${\bf\sigma^E}$ is  the elastic stress tensor and ${\bf\sigma^V}$ is
the viscous stress tensor. The former is related
to the layer displacement $u$
and can be obtained from the following equation relating the change of the
free energy $f_b$ to the displacement:
$$df_b=\sigma^E_{zj}d\left(
{{\partial u}\over{\partial r_j}}\right),\eqno(4.4)$$
Since u is the  z-component of the displacement vector, the only nonvanishing
component of this tensor have (z,j) indexes.
Expanding (3.20) and equating the expansion to the right hand side of (4.4)
we find
$$\sigma^E_{zj}={{\partial f_b}\over{\partial u_{,j}}}-
{{\partial}\over{\partial r_i }}
{{\partial f_b}\over{\partial u_{,ij}}}.\eqno(4.5)$$
Here $u_{,j}$ and $u_{,ij}$ are the first and the second derivatives of $u$
with respect to the  components of ${\bf r}$.
More explicitely we find:
$$\sigma^E_{zz}=B{{\partial u}\over{\partial z}},\eqno(4.6)$$
$$\sigma^E_{zx}=-K{{\partial}\over{\partial x}}\left(\triangle_\bot u\right)
,\eqno(4.7)$$
$$\sigma^E_{zy}=-K{{\partial}\over{\partial  y}}\left(\triangle_\bot u\right).
\eqno(4.8)$$
All other terms are zero. Note that elastic stress tensor is not symmetrical
due to the curvature elasticity and the associated surface torque
${\bf C}\hat{\bf n}$ . The nonvanishing components of the tensor ${\bf C}$
are
$$C_{xy}={{\partial f_b}\over{\partial u_{,yy}}},\eqno(4.9)$$
and
$$C_{yx}=-{{\partial f_b}\over{\partial u_{,xx}}}.\eqno(4.10)$$
The elastic stress tensor satisfies the torque balance equation
$$C_{ij,j}-\epsilon_{ijk}\sigma^E_{jk}=0\eqno(4.11)$$
automatically  (we assume as stated in section III that $\hat{\bf n}$ is
normal to layers).

In order to calculate the expression for the viscous stress tensor we have to
find the thermodynamic fluxes and forces
associated with the entropy production. One finds for the isothermal processes
the following expression for the change of the entropy $s$ in time:
$$T{{D s}\over{D t}}=\left({{D u}\over{D t}}-v_z\right)G+\sigma^V_{ij}A_{ij},
\eqno(4.12)$$
where
$$G=\sigma^E_{zj,j}\eqno(4.13)$$
is the elastic force normal to layers and
$$A_{ij}={1\over 2}\left({{\partial v_i}\over{\partial r_j}}+
{{\partial v_j}\over{\partial r_i}}\right).\eqno(4.14)$$
Now from the Onsager relations between thermodynamic
forces ($G,\sigma^V_{ij}$) and fluxes ($A_{ij}, (Du/Dt-v_z)$)
we find
$$\eqalign{
\sigma_{ij}^V=&\mu_1\delta_{ij}A_{kk}+\mu_2\delta_{iz}\delta_{jz}
A_{zz}+\mu_3A_{ij}\cr+&\mu_4\left(\delta_{iz}A_{zj}+\delta_{jz}A_{zi}\right)+
\mu_5(\delta_{iz}\delta_{jz}A_{kk}+\delta_{ij}A_{zz})\cr}
\eqno(4.15)$$
and
$${{Du}\over{Dt}}-v_z=\lambda_pG.\eqno(4.16)$$
Here $\mu_i$ ($i=1\cdots 5$) are independent viscosities.
In  the following we shall take $\sigma_{ij}^V=\mu A_{ij}$ for
simplicity.
Eq.(4.16) is formally equivalent to the Darcy law in porous medium.
In our case the lamellar phase plays both the role of the fluid
and of the porous medium. It can be shown that the permeation
coefficient $\lambda_p$,
$$\lambda_p\sim {{D_\parallel v_{\rm mol}}\over{k_BT}}\eqno(4.17)$$
where $D_\parallel$ is the diffusion coefficient normal to the layers
and $v_{{\rm mol}}$ is a molecular volume. The permeation coefficient is much
smaller in lyotropic (e.g. amphiphilic systems) than in
thermotropic smectic liquid crystals. This coefficient can be estimated
from the measurements of the edge dislocation mobility as will be shown in
Section XIII.

\centerline{\bf V. Static bulk properties of dislocations}

Let us consider an elementary edge dislocation (of unit Burgers vector)
located along the y-axis at $x=l$ and $z=h$.
The displacement $u$ is a multivalued function in the $z=h$ plane i.e.
$$u(x,z)=\cases{0, &if $x\le l$;\cr {\rm sgn}(z-h)d/2,&if
$x>l$.\cr}\eqno(5.1)$$
This is equivalent to the condition
$$\oint{{\hat{\bf n}}\over d}d{\bf l}=1,\eqno(5.2)$$
where the contour of integration is around the dislocation line.
The distortions induced by a dislocation are described
by the following equation$^{7}$:
$${{\partial^2 u}\over{\partial z^2}}-\lambda^2\triangle_\bot^2 u
=0\eqno(5.3)$$
obtained from the minimization of the total bulk distortion
energy $F_b$,
$$F_b=\int d{\bf r} f_b.\eqno(5.4)$$
Here $f_b$ is the energy density given by Eq.(3.20) and $\lambda=\sqrt{K/B}$.
The equilibrium solution of Eq.(5.3) satisfying condition (5.1)
is as follows$^7$:
$$\eqalign{u_b(x,z)=&{\rm sgn}(z-h)\left({d\over{4\pi}}
\int dq \exp{(-\lambda q^2\vert z-h\vert)}{{\exp{(iq(x-l))}}\over{i(q-i0^+)}}
\right)\cr=&{d\over4}{\rm sgn (z-h)}\left(1+
{\rm erf}\left({{x-l}\over{2\sqrt{\lambda\vert z-h\vert}}}\right)\right)\cr},
\eqno(5.5)$$
where ${\rm erf}(t)=2/\sqrt{\pi}\int_0^{t} \exp (-u^2) du$
is the error  function.
The solution is valid outside the dislocation core of size $r_c$.
For dislocations of Burgers vector, ${\bf b}$,
of length $nd$, for which
$$\oint{{\hat{\bf n}}\over d}d{\bf l}=\pm n,\eqno(5.6)$$
the solution is simply given by $\pm n u_b$ (here $n$ is a
positive integer).
Inserting the solution (5.5) into $F_b$ (Eq.(5.4)) and introducing the cutoff
$r_c$ in (5.4) we find the energy of the edge dislocation line per unit length
$F_b/L_y$:
$$E_0=E_c+{{\sqrt{KB} (nd)^2}\over {2r_c}},\eqno(5.7)$$
where $E_c$ is the core energy. As expected this energy
does not depend on $h$ and $l$
in the infinite systems, but as we shall see in the
next section it does depend
on the location of dislocation $(l,h)$
in the finite system bounded by surfaces.
The energy of dislocation per unit length is finite in lamellar phases
whereas in solids it
diverges logarithmically with the size of the system.
Also stresses do not vary like $1/r$ as in solids. We find from
section IV (Eq(4.6)) and Eq(5.5) that in lamellar phases
$$\sigma^E_{zz}=-{{(x-l)d}\over{8\vert z-h\vert\sqrt{\pi\lambda \vert z-h\vert
}}}
\exp{\left(-(x-l)^2/(4\lambda\vert z-h\vert)\right)}.\eqno(5.8)$$
Thus $\sigma^E_{zz}$ is large only for $4\lambda\vert z-h\vert\ge x^2$.
Outside this region the stress vanishes very quickly because of the
fluidity of
layers.
Let us now discuss the core structure of an edge dislocation.
Let us assume that the core is nematic. We also assume that the core is
anisotropic i.e. along the z direction it has the size $2 r_c$ while along the
x direction it has negligible size. Then
$$E_c=2\gamma r_c\eqno(5.9)$$
where $\gamma$ is  roughly proportional to the nematic-smectic
surface tension.
If we now minimize $E_0$ (Eq(5.7) with respect to $r_c$ we find
$$E_0=2\sqrt{B\lambda\gamma}nd\eqno(5.10)$$
and
$$r_c={{nd}\over 2}\sqrt{{{B\lambda}\over{\gamma}}}.\eqno(5.11)$$
When $\gamma$ is small the core size can be very large as shown in (Fig.4b).
The total energy is now proportional to the length of the Burgers
vector, $nd$, of
an  edge dislocation and this explains why dislocations may group
together. When their Burgers vector is large (giant dislocations
with $n\ge 10$), the
core of dislocation can split$^{17,11,7}$
into two $\pm 1/2$ disclinations a distance $nd/2$ apart (Fig.4c)
The core energy is in this case,
$$E_c={{\pi K}\over 2}\ln{\left({{nd}\over{2r_0}}\right)}+w_c,\eqno(5.12)$$
where $w_c$ is the disclination core energy and $r_0$ is the
disclination core size.
Now the core energy is lowered by
the gathering of elementary edge dislocations
into dislocations of large $n$. The core structure of disclinations in
smectics is not known, although recently the disclination
core structure in nematic liquid crystals
has been thoroughly studied in the
Landau-de Gennes model$^{38}$ and
in computer simulations for hard rod systems$^{39}$.
The core structure of this defect depends on the length of molecules.
For short molecules, the core is
biaxial with molecules perpendicular to the
disclination line, whereas for long molecules they are parallel to this line.
The core size is proportional to molecular thickness, rather than
molecular length.

Let us now discuss the screw dislocation properties.  In this
case (Fig.4d) the Burgers vector is parallel to the dislocation
line and perpendicular to layers.  Recent discovery of the TGB
(Twist Grain Boundary) phase consisting of lattice of screw
dislocations$^{40,41}$ (analogous to the type II
superconductors) brought the renewal interests in these
defects$^{42}$.  The static properties of screw dislocations
have been intensively studied$^{42-47}$ and here we shall
briefly present the major results of this properties.  The
distortions induced by screw dislocation of strength $n$ located
along the $z$ axis (perpendicular to smectic layers) are also
described by Eq.(5.3) and the condition(5.6). For small
distortions we have: $$\oint{\nabla_\bot u}d{\bf l}=\pm
nd,\eqno(5.13)$$ Due to the symmetry $u$ does not depend on $z$
and the layer thickness $d$ is equal to the undistorted value
$d_0$.  We find in this case the equilibrium solution of the
form$^{43}$: $$u_b(x,y)=\pm {{nd\phi}\over {2\pi}}=
\mp{{nd}\over {2\pi}}\left(\arctan{\left({y\over x}\right)}-\pi+
(\pi/2){\rm sgn}(y)\right).\eqno(5.14)$$
where $\phi$ is the polar angle in cylindrical coordinates.
The solution is only valid far away from the core, i.e. for $\rho\gg r_c$,
where $r_c$ is the core size and $\rho$ is the distance, in polar
coordinates,
form the screw axis
(given by $z=0$).
The total distortion energy is the core energy only, since the strain energy
is zero. In solids the strain energy
diverges logarithmically with the size of the system.
If we include anharmonic terms in the distortion energy
(section III,  Eq(3.15)) and the splay term, which for distances not too far
from the core can compete with the compressional term, we find the energy per
unit
length of the screw dislocation$^{43,46}$ in the following form:
$$E_0=E_c+{{(nd)^4B}\over {128\pi^3r_c^2}}+{{(nd)^4K_3}\over{64\pi^3 r_c^4}}.
\eqno(5.15)$$
It should be noted that we have used the solution (Eq.(5.14)) which
followed from the harmonic approximation (Eq.(5.3)).
Since $K_3$ and $B d^2$ are of the same order of magnitude, so both
contribute significantly to the distortion energy.
Note however that this elastic energy is three orders of magnitude
smaller then the energy of the edge dislocation. The core energy
(core size $r_c$) can be roughly estimated$^{44}$ as:
$$E_c=k_B(T_{NA}-T)\pi r_c^2/v_{\rm mol},\eqno(5.16)$$
where $v_{\rm mol}$ is the molar volume and $T_{NA}$ is the
transition temperature from the nematic liquid to the lamellar
phase. The core energy vanishes at the transition which means that the
formula (5.16) is valid only for the second order phase transition.
In the case of the first order phase transition we find:
$$E_c=\pi r_c^2 {{\Delta H (T_{NA}-T)}\over {T_{NA}}}+2r_c\gamma,\eqno(5.17)$$
where $\Delta H$ is the latent heat (per  unit volume)
of the transition and $\gamma$ is the
nematic-smectic surface tension. Here it has been assumed that the
core is filled with undercooled nematic phase. Similarly as we did in the case
of edge dislocations (Eqs(5.10-11)) we could minimize $E_0$ with respect
to $r_c$. This calculation shows that the energy of the screw
dislocation increases faster than the length of the Burgers vector $nd$
so that elementary screw dislocations are favored.
In fact the exact core structure of screw dislocation
is not known in general, although some preliminary studies have been
done for smectic liquid crystals$^{42,46}$.
These studies indicate that the core is nematic
(although there is a possibility for the completely isotropic core).
Since layers are not well defined inside the core
one has to use there
the Landau-de Gennes description in terms of
the order parameters$^{7}$. The smectic order parameter $\epsilon$, the
nematic
order parameter $s$ and the local angle between the director field
and the $z$ axis, $\vartheta$,  are shown in Fig.14. The nematic order
parameter
does not change in comparison to its bulk value $s_b$. The smectic order
parameter
decreases to zero in the core as $\rho^n$, and approaches the bulk value
$\epsilon_b$ as $-1/\rho^4$. Within the core the director $\bf n$ tends to
lie along
the screw axis and $\vartheta\sim\rho$ close to the center of the core.
The director ${\bf n}={\bf e_\rho}\cos\alpha_b\sin\vartheta+{\bf e_\phi}
\sin\alpha_b\sin\vartheta+{\bf e_z}\cos\vartheta$ should be identified
far from the core with the layer normal. Far from the core one finds
$\alpha_b=\pi/2$ and $\vartheta =\arctan (nd/2\pi\rho)$.
The size of the core increases with the strength of the screw dislocation
$n$.
Finally we note that the internal stress
inside the core (giving rise to radial force)
is balanced in
the smectic by the non-uniform pressure $p=p_0+
(nd)^4B/128 \pi^4\rho^4$, which can be as large as twice the athmospheric
pressure, $p_0$, of a defect free sample.
The case of the isotropic screw core structure has also been
studied$^{42,46}$.
These cores are energetically unfavorable, although they may be stabilized
close
to the smectic liquid crystal --
isotropic phase transition. The core structure for the polymers or amphiphilic
systems certainly cannot be nematic since the nematic phase is
not present in these systems. To our knowledge these cores have not been
studied in detail (see however Refs.11,44).

Finally let us mention  an important property of screw dislocations, namely
the fact that although their energy per unit length
is small their line tension is
large$^{44}$. In order to calculate the line tension of the screw dislocations
let us
assume that the dislocation is tilted with respect to the layers, making
an angle $\theta$ with the z axis Fig.15. For very small $\theta$
the displacement is given by:
$$u_b(x,y,z)={{nd}\over{2\pi}}\arctan{{y-\theta z}\over{x}}\eqno(5.18)$$
and the total energy is
$$E_0(\theta)=E_0(\theta=0)+{{(nd)^2B}\over{4\pi}}\theta^2\ln{{R\over r_c}},
\eqno(5.19)$$
where $R$ is the size of the system or the distance between the neighbouring
screw dislocations of opposite signs. The line tension is defined as follows:
$$T_{{\rm screw}}=E_0(\theta=0)+{{d^2E_0(\theta)}\over{d\theta^2}}
\Big\vert_{\theta=0}\eqno(5.20)$$
and it follows from (5.19) that $T_{{\rm screw}}\gg E_0$ contrary to
the case of edge dislocations for which $T_{{\rm edge}}\sim E_0$.
The main difference between the two types of dislocations is that
edge dislocations located along the y axis
do not change their character when tilted in the x-y plane
with respect to  the y-axis,
while the screw dislocations along the z axis
acquire the edge character when tilted
with respect to the z axis. This result explains why screw dislocations
prefer to be perpendicular to smectic layers and
do not tilt easily (Fig.16). We shall also see that this property is
crucial for
understanding the helical instability under compression or dilation
normal to layers.

\centerline{\bf VI. Equilibrium location of dislocations in finite systems}

Experimental systems are always bounded by surfaces. In Meyer et al$^{13}$
experiment two glass plates bound the lamellar phase (Fig.5).
Thin freely suspended smectic liquid crystal films$^{23}$ (Fig.8)
are attached to the aperture and bounded from above and below by
a smectic free surface (smectic-air interface).
Smectic droplets$^{29,30}$
deposited on a solid substrate (Fig.10)
are bounded by the solid surface and
the free surface. The diblock copolymer lamellar film deposited on a solid
substrate shown in Fig.7 is also bounded by a free surface.
In the finite system dislocations are influenced by the bounding surfaces.
The boundary effects can be incorporated in the form of the surface distortion
energy. It should be noted that contrary to the solids, the surface energy of
distortion in lamellar liquid phases is comparable to the bulk energy.
In solids the energy scales for the surface and the bulk are well separated
with surface energy being much smaller than the bulk energy.

The influence of surfaces on dislocations has been
studied by Pershan$^{19}$. However his analysis
was not sufficient to explain interactions with a free surface$^{25}$.
The problem has been first analyzed theoretically by Lejcek and Oswald$^{20}$
in the framework of the image dislocation approach.
In the semi-infinite system bounded by the free surface at $z=0$
the total energy contains the bulk part (5.4) and the surface
part$^{20}$:
$$F_s={1\over 2}\int d{\bf r}_\bot\gamma\vert\nabla u_s\vert^2,\eqno(6.1)$$
where $\gamma$ is the smectic-air surface tension and
$u_s=u({\bf r}_\bot,z=0)$ is the surface displacement. Minimizing
$F_b+F_s$ gives Eq.(5.3) and a boundary condition at the free surface:
$$B{{\partial u}\over{\partial z}}=\gamma\triangle_\bot u_s.\eqno(6.2)$$
Let us now consider a simple case of a dislocation
of Burgers vector $b=nd$ located
at $z=h$. It can be shown that both equation (5.3) and
(6.2) are satisfied by the solution (5.5)
if we introduce a symmetric image dislocation
(located at $z=-h$) of
Burgers vector:
$$b_i=b{{\gamma-\sqrt{KB}}\over{\gamma+\sqrt{KB}}}.\eqno(6.3)$$
and take the linear combination of the bulk displacement field ($u_b$
Eq(5.5))
for both dislocations $u=n u_b(h)+n_i u_b(-h)$.
Note that $b_i=n_i d$
does not have to be a multiple of $d$ since it is a virtual
dislocation and consequently $n_i$ in general is not an integer.
We shall see in section VIII that two dislocations
repell each other when have the same sign and attract each other for
opposite signs. Thus it follows that the surface attracts dislocation
for $\gamma <\sqrt{KB}$ and repels it in the opposite case.

These calculations can be generalized to more complicated situations,
when for example
we have two surfaces$^{20,30}$, but in this case we have to introduce
an infinite number of image dislocations, which is not very convenient.
For this reason we shall adopt another
approach to  the problem based on de Gennes method of calculation$^{7}$.
We also note that the surface energy considered by Lejcek and Oswald is
incomplete since it does  not contain the surface curvature term
which influences very strongly the equilibrium location
of dislocations in the case of small surface tension$^{48,49}$.
The surface distortion energy  with the curvature term
has the following form$^{50}$
$$F_s={1\over 2}\int d{\bf r}_\bot\left(\gamma\vert\nabla u_s
\vert^2+
K_s(\triangle_\bot
u_s)^2\right).\eqno(6.4)$$
Here the new quantity is
the surface elastic constant, $K_s$. It should be identified
with the bending elastic constant of the last lamellar layer
(at the surface). In general the properties of the surface layer are different
from the bulk layers and
the value of the surface constant can be different from the bulk
value. In particular, if the surface layer is the same as the bulk one,
they are related by the formula: $K_s=Kd$, where $d$ is the layer spacing.
The fact that the surface bending  elastic constant must be included follows
from the comparison of the distortion energy in the discrete$^{51}$ and
continuous$^{50,52}$ representation.
In the discrete representation the layers have indexes $i=0,1,2...$
and $u(x,z)=u_i(x)$. Then the last term in Eq(6.4) appears naturally with
$K_s=Kd$. In the discrete representations we sum over $i$ instead
of integrating over $z$ and consequently
each layer, including the surface one, has the bending
elastic constant equal $Kd$.

Let us consider the experimental system
shown in Fig.7 with the substrate located at $z=D$ and an
elementary edge
dislocation ($n=1$) located at $z=h$ and $x=l$.
The condition for the edge dislocation is given by Eq.(5.1).
Now minimizing the total energy $F=F_b+F_s$ (Eq(5.4) and (6.4))
we find
Eq.(5.3) plus the boundary condition at the free surface ($z=0$):
$$-B{{\partial u}\over{\partial z}}-\gamma\triangle_\bot u_s
+K_s\triangle^2_\bot u_s=0\eqno(6.5)$$
and the boundary condition at a solid substrate,
$$u(x,z=D)=0.\eqno(6.6)$$
The equilibrium solution satisfying
Eqs(5.1,5.3,6.5,6.6) is given by $u_{eq}(x,z)$,
$$u_{eq}(x,z)=u_b(x,z)+u_p(x,z)\eqno(6.7)$$
where the singular bulk part of the solution, $u_b$, is given
by Eq.(5.5)
and the particular non-singular part of the solution, necessary to satisfy
the boundary conditions (6.5-6), is $^{48}$
$$u_p(x,z)={d\over{4\pi}}\int dq{{\exp{(iq(x-l))}}\over{iq}}
\left(\exp{(-\lambda q^2 z)}f_1(q)+\exp{(\lambda q^2 z)}f_2(q)\right).
\eqno(6.8)$$
Here
$$f_1(q)=-{{a_1(q)e^{-\lambda q^2h}+
a_1(q)e^{\lambda q^2(h-2D)}}\over
{a_2(q)+a_1(q)e^{-2\lambda q^2 D}}}
,\eqno(6.9)$$
$$f_2(q)={{a_1(q) e^{-\lambda q^2(h+2D)}-
a_2(q)e^{\lambda q^2(h-2D)}}\over
{a_2(q)+a_1(q)e^{-2\lambda q^2 D}}}
\eqno(6.10)$$
and
$$a_1(q)=1-\alpha_s-\lambda_s^2q^2,\eqno(6.11)$$
$$a_2(q)=1+\alpha_s+\lambda_s^2q^2.\eqno(6.12)$$
Here $\alpha_s=\gamma /\sqrt{KB}$ is the dimensionless constant,
whereas $\lambda=\sqrt{K/B}$ and $\lambda_s=\sqrt{K_s/\sqrt{KB}}$
are two microscopic lengths.
The total distortion energy $F=F_b+F_s$
can be written in the following convenient form:
$$F(h)=E_0+{1\over 2}B\int dx{{\partial u_p(x,z)}\over{\partial z}}
\left(u_b(x,z=h^{-})-u_b(x,z=h^{+})\right).\eqno(6.13)$$
Inserting $u_b$ and $u_p$ into Eq(6.13)
gives $F(h)$ as a function of $h$ in the
following form:
$$F(h)=E_0+{{\sqrt{KB} d^2}\over{8\pi}}\int
dq{{-a_1(q)e^{-2\lambda q^2h}+ a_2(q)e^{2\lambda q^2(h-D)}-2a_1(q)e^{-2\lambda
q^2D}}\over {a_2(q)+a_1(q)e^{-2\lambda q^2 D}}}.\eqno(6.14)$$
As one notices the equilibrium location of the dislocation is
still to be determined
The equilibrium
location of the dislocation, $h_{eq}$, corresponds to the minimum of $F(h)$.
$${{\partial F(h)}\over{\partial h}}=0.\eqno(6.15)$$
First let us consider the case of the infinite thickness of the film
$D=\infty$. The energy as a function of $h$ is shown in Fig.17.
For $\alpha_s=0$ we find $h_{eq}=1.08 K_s/(2 K)$ and for
$\alpha_s\le 0.2$ we get roughly
$h_{eq}\ge (K_s/(2 K))((1+\alpha_s)/(1-\alpha_s))$.
For larger values of $\alpha_s$, we find that the following
inequality holds:
$${{1+\alpha_s}\over{1-\alpha_s}}>{{2 h_{eq}K}\over{K_s}}> {1
\over{1-\alpha_s}}
.\eqno(6.16)$$
For small surface tension the dislocation is stabilized at
a finite distance due to the surface bending elastic constant $K_s$.
Finally for $\alpha_s\to 1$ we have $h_{eq}\to\infty$,
which means that the dislocation is no longer stabilized at a finite distance
from the
interface. This result is the same as obtained within the image
dislocation approach of Lejcek and Oswald$^{20}$.
For large surface tension $\alpha_s\gg 1$ and finite but large $D$
the dislocation is stabilized
roughly in the middle of the system at $h_{eq}\approx D/2$$^{19,20,22,48}$.
For small surface tension and finite $D$ the location of dislocation in a
system depends strongly on $D$. (Fig.18)
For small $K_s$ the dislocation is
close to the interface, while for larger $K_s$ it
moves towards the center of the system. For thin films there is a strong
deviations from the linear dependence of $h_{eq}$ on $K_s$.
This is of course due to
the strong repulsion of the dislocation from the solid substrate,
which cannot be compensated by any finite value of $K_s$.
Finally dislocations in free standing film
shown in Fig.8a,
are located exactly in the middle of the film,
since $\gamma\gg\sqrt{KB}$ in this case$^{53}$ and the film is symmetric.

Usually the surface tension for the air-lamellar interface is
about $\gamma=30$ dyn/cm (this value
characteristic for air-organic liquid interface (free surface)) and
in general larger by an order of magnitude than $\sqrt{KB}$.
However, there are also many systems characterized by a small
value of the surface
tension (smaller than $\sqrt{KB}$).
They include nematic-smectic interface (for liquid crystals), microemulsion-
lamellar interface in amphiphilic systems and AB diblock copolymer
lamellar-A (or B) homopolymer melt interface.
Our results can be used to describe
dislocations in all these systems.

Here we may also
ask the question concerning the average distance, $l$, between
dislocations in a smectic meniscus (Fig.8a).
Neglecting the elastic interactions
we find
$$l={{\gamma d\exp{(x/\xi_0)}}\over{\gamma_{AW}-\gamma_{LW}-F(h)/d}},\eqno(
6.17)$$
for large film thickness and far from the aperture ($x\geq\xi_0$).
Here $\gamma_{AW}$ is the air-aperture
surface energy(per unit area), $\gamma_{LW}$ is the
lamellar-aperture surface energy and $F(h)$ is the energy of dislocation
(per unit length) and $h=D_0/2$ is half of the thickness
of the meniscus close to the aperture.
$\xi_0=\sqrt{\gamma/\rho g}$ is the capillary
length, $\rho$ is the density and $g$ is the gravitational accelaration.
This length can be very large (of the order of millimeters).
Close to the aperture the dislocations are very close, interact strongly
and the formula (6.17) may not be valid.
That is why the smectic meniscus
offers a very good opportunity to study the competition between gravitational
forces and elastic interactions between dislocations.
The system can also develop sharp step defects at the surface
close to the aperture. These steps do not have to be related to dislocations
inside the meniscus (see discussion on steps in Sec.II and
droplets in Fig.10ab).
The meniscus also plays an important role for explaining the
tendency of the film to get thinner.
Indeed, the pressure inside the film $P$ is less than the athmospheric
pressure, $P_0$, because the meniscus is curved. The
equilibrium condition at the surface gives $P-\sigma_{zz}^E=P_0$ and
thus  the layers are compressed. Consequently the film tends to get thinner.
Let us assume that we have nucleated a dislocation loop, say by
suddenly pulling a film. Then
we find that
the dislocation loop of size $R$ and Burgers vector of length $nd$
will spontaneously grow for $R>R_c=F(D/2)/((P_0-P) nd)$. Here $D$
is the thickness of the film.

Finally let us discuss the clustering of like dislocations (Fig.19ab)
in a wedge shaped samples$^{13, 26, 54, 55}$.
Calculation of the energy in the case shown  in
(hatched region in Fig.19a)
gives, for dislocations of Burgers vector $nd$:
$$E=2 E_0(n)+{{B(nd)^3}\over{12 \alpha D}},\eqno(6.18)$$
where $\alpha$ is the wedge angle, and $E_0(n) $
is given by Eq(5.7) and $D$
is the average thickness
of the sample over a distance between the neighbouring
dislocations.
For the case shown in Fig.(19b) when dislocations are grouped in pairs of
the doubled Burgers vector $2nd$ we find:
$$E=E_0(2n)+{{B(nd)^3}\over{3\alpha D}},\eqno(6.19)$$
where $E_0(2n)$ is given by (5.7).
If $2E_0(n)>E(2n)$ then we find that dislocations form pairs
for the angle, $\alpha$, larger then the critical angle $\alpha_c$,
$$\alpha_c={{B(nd)^3}\over{4D(2E_0(n)-E_0(2n))}.}\eqno(6.20)$$
Experimentally$^{54}$ one finds $\alpha_c\sim 10^{-3}$ when $D=100\mu$m
and $n=1$.
Then from (6.20) we find $2E_0(n)-E_0(2n)\sim 10^{-3}K$, which means
that $2E_0(n)\sim E_0(2n)$. Although they are not equal exactly as
predicted by (5.10) they are very close, indeed. Certainly the small
differences can come from the subtle changes in the core structure and/or
anharmonic terms. Similar studies have been done for ferrosmectics$^{56}$.

\centerline{\bf VII. Edge profiles}

Inserting $h_{eq}$ into
$u_{eq}$ we get equilibrium distortions in the film; the edge profile at the
interface is given by $u_{eq}(x,z=0)$ (see Fig.7).
The influence of the surface tension on the width of the
profile is shown
in Fig.20. Large surface tension broadens the edge
profile$^{22,48}$ and
for a symmetric film of size $D$ the
width of the profile, $\Delta x$ scales as $^{22}$
$\sqrt{\gamma D/B}$. Theoretical analysis$^{22,48}$
explained qualitatively and quantitatively the results of
experiments$^{21,22}$ performed on thin diblock copolymer
films deposited on solid substrate.
The comparison between the theory and experiment$^{23}$ is shown in Fig.21.
The values of parameters obtained were as follows: $\gamma =23.5$ dyn/cm$^2$
(see also Ref.57), $\lambda=\sqrt{K/B}\approx
d=315$\AA, $\sqrt{KB}=0.9$ dyn/cm,
$\gamma /\sqrt{KB}\approx 26$. Also
the scaling law for the width i.e. $\Delta x\sim\sqrt{D}$ has been
verified experimentally. The same analysis has been also
successfully applied to the analysis of the edge profiles in
the smectic phase formed by
main-side chain liquid crystal polymers deposited on a solid
substrate$^{58,59}$. In this experiment, the x-ray scattering has
revealed the
unsually large roughness at the surface$^{58}$.
Then the scanning tunneling microscopy has shown the clear sign of steps
at the free surface, an indication of dislocations inside the sample$^{59}$.
The value of the parameters obtained in this  experiments are as follows:
$\gamma/\sqrt{KB}=5$, $\lambda=\sqrt{K/B}=101$\AA$\approx 5d$.
The smectic length $\lambda$ is much larger than the smectic period $d$,
reflecting the fact
that  in the main-side chain liquid crystalline polymer we have two
characteristic lengths. One  is the size of the mesogenic unit
related to $d$ and one is the full length of the polymer chain related to
$\lambda$. We also see that  $K$ increases with the length of the polymer.

So far we have assumed that steps at the free surface are
due to dislocations inside the sample.
However, it should be noted that steps at the surface do not have to
be related to dislocations(Fig.22) and can have the
same origin as steps in solids.
Theoretically, the energy associated with a step$^{60}$
(Fig.14) is an order of magnitude
larger than the one associated with the dislocation (Fig.7).
The extra cost in the distortion energy per unit length
shown in Fig.22. is
$\Delta\gamma d/2$, where $\Delta\gamma$ is the surface tension difference
for the PS and PBMA block of copolymer. It is roughly equal to $5 d$ dyn/cm.
At the same time the extra energy associated with dislocation is
$\sqrt{KB}d/2$
which is equal to $0.5 d$ dyn/cm. An order of magnitude difference
clearly shows that the configuration with dislocations (Fig.7)
is favored over the configuration with steps(Fig.22) in
this particular system.
Also TEM micrographs$^{22}$ clearly show that under the step at the surface
we have indeed edge dislocations.
However as we have discussed in Section II
in some cases the configuration with steps can have lower energy than
the configuration with dislocations.

For small surface tension the edge profiles$^{49}$
strongly depends on $K_s$.
Fig.23 shows the results for $\gamma=0$.
The most interesting feature of $u_{eq}(x,z=0)$ is the nonmonotonic behavior
of the profile as a function of $x$, observed for finite $D$.
This is due to the finite size of the
film and zero surface tension, since only in this case it is
favorable to distort more the interface than the bulk.
Large surface tension
would make the profile perfectly monotonic. In very thin films
we expect that the surface tension is important in determining
the step structure when
$\gamma\approx\sqrt{K_s B/D}$. In thick films the surface tension
dominates when $\gamma\ge\sqrt{KB}$.
We find that the width of
the profile grows  with increasing $K_s$.

\centerline{\bf VIII. Elastic interactions between dislocations}

So far we have considered the properties of a single
dislocation inside the system. Now let us consider two edge dislocations
in the infinite system:
one located at $(x,z)=(0,0)$ and one at $(x,z)=(l,h)$.
Since the equations for the displacement $u$ are linear the
total displacement, $u_b$ due to dislocations 1 and 2 is the
linear combination of individual displacements $u_b^{(1)}$ and
$u_b^{(2)}$. The distortion energy
can be conveniently written (using Eq.(5.3))
in the following form:
$$\eqalign{F_b(h,l)=&E_0^{(1)}+E_0^{(2)}
+{1\over 2}B\int dx{{\partial u_b^{(2)}(x,z)}\over{\partial z}}
\left(u_b^{(1)}(x,z=0^{-})-u_b^{(1)}(x,z=0^{+})\right)\cr &
+{1\over 2}B\int dx{{\partial u_b^{(1)}(x,z)}\over{\partial z}}
\left(u_b^{(2)}(x,z=h^{-})-u_b^{(2)}(x,z=h^{+})\right)\cr}.\eqno(8.1)$$
Inserting the solutions obtained in section V we find$^{10,19} $
$$F_b(h,l)=E_0^{(1)}+E_0^{(2)}
\pm{{\sqrt{KB} n^{(1)}n^{(2)}
d^2}\over {4\pi}}\int dq\exp{(-\lambda q^2 h)}\cos{(ql)}
\eqno(8.2)$$
Here plus sign corresponds to like dislocations (same sign
of the Burgers vector i.e. same sign in condition
(5.6))
and minus sign to the unlike dislocations; $n^{(i)}$ $i=1,2$ is the
strength of the dislocations (see Eq(5.6)).
The interaction energy is negligible for
two edge dislocations located in the same
plane ($z=h=0$). Also for $l^2\gg 4\lambda h$ the interactions are weak.
Two like dislocations (same Burgers vector) always repel each other along
the x direction ($\partial F_b/\partial l<0$) and attract each other along
the $z$ direction only for $l^2>2\lambda h$ ($\partial F_b/\partial h>0$). For
$l^2<2\lambda h$  two like dislocations repel each other also along the
z direction.

For two screw dislocations located at $x=0$ and $x=l$ one finds$^{43}$
that the interaction energy is zero in the harmonic approximation.
If one includes anharmonic terms$^{46}$ in a simple perturbation
scheme one finds that two screw dislocations interact with a logarithmic
potential (per unit length of dislocation line):
$$F_b(l)=\pm 2 \left(B{{n^{(1)}n^{(2)} d^2}\over{4\pi^2}}+K_3
\left({{n^{(1)}}\over{
n^{(2)}}}+{{n^{(2)}}\over{n^{(1)}}}\right)\right)\ln{(l/r_0)},\eqno(8.3)$$
where $r_0$ is the larger core size of two screw dislocations.
According to Eq(8.3) two screw dislocations of same sign
(plus sign in Eq(8.3)) attract  each other with
the long range potential. Two screw dislocations of
opposite sign (minus sign in Eq.(8.3))
repel each other preventing their annihilation.
Although the procedure applied to this problem$^{46}$ casts some
doubts on the full validity of this equation,
nonetheless it shows that the
complete role of the anharmonic terms in the theory of
dislocations is not known.

So far we have considered the infinite system with no boundary effects.
Here we go one step further and consider the finite
symmetric system, bounded by two free surfaces (Fig.8a).
The distortion energy is given by the bulk term (Eq.(5.4)) and
two surface terms
$F_s$ (see Eq.(6.4)). Minimizing this energy we find Eq.(5.3) and
two boundary conditions for each free surface (see Eq(6.5)).
The solution is the sum of individual solutions
(since the equations are linear). Because in our system
$\gamma>\sqrt{KB}$ two edge dislocations are located in the
same plane in the middle of the film and thus their bulk interactions
are negligible.
Now the only elastic interactions are surface induced. Inserting the
equilibrium solution for two
dislocations
at a distance $l$ apart we find
the surface induced elastic interaction per unit length of dislocation:
$$F_s(l)=-{{\sqrt{KB}n^{(1)}n^{(2)}d^2}\over{(2\pi)^2}}\int dq
f(q)(1\pm e^{iql})\eqno(8.4)$$
where the plus sign corresponds to like dislocations and
minus to unlike dislocations.
$$f(q)={{a_1(q)}\over{a_2(q)\exp{(q^2\lambda
D)}+a_1(q)}}$$
and $a_1(q)$ and $a_2(q)$ are given by Eqs.(6.11-12).
The total distortion energy  for these dislocations is given by the
sum of $F_s(l)$ and $F_b(h=0,l)$ (Eq(8.4),Eq.(8.2)).
For large $\alpha_s$ (how large $\alpha_s$ should be depends on the film
thickness and the surface elastic constant, typically $\alpha_s>2$ is
sufficient)
the interactions are repulsive for all
distances for like dislocations and attractive for opposite
dislocations. For $\alpha_s\gg 1$ we find from Eq.(8.4) the following
form of the elastic interaction energy:
$$F_s(l)\approx
{{\alpha_s\sqrt{KB}n^{(1)}n^{(2)}d^2}\over{16\pi\xi}}
(1\pm\exp{(-l/\xi)}),\eqno(8.5)$$
where $\xi =\sqrt{\alpha_s D\lambda /2}=\sqrt{\gamma D/2B}$
is proportional to the size of the edge profile as could be expected.
We have neglected the bulk contribution to
the interaction energy since at large $l$ it
decays as $\exp{(-(l/d)^2)}$.
In the extreme limit of $\alpha_s\to\infty$
(which correspond to the system sandwiched between two
walls) we find that this energy diverges for like dislocations.
For large $\alpha_s$ we find for like dislocations:
$$F_s(l)={{\alpha_s\sqrt{KB}n^{(1)}n^{(2)}d^2}\over{8\pi\xi}}-
Bn^{(1)}n^{(2)}d^2l/(8\pi D).\eqno(8.6)$$
and for unlike ones:
$$F_s(l)=Bn^{(1)}n^{(2)}d^2l/(8\pi D).\eqno(8.7)$$
As could be expected unlike dislocations attract each other while
like dislocations repel each other in the system with
large surface tension.

For small $\alpha_s$, but still larger than 1,
the behavior of
$F_s(l)$ is more complex, than in the previous case.
For like dislocations
the potential is repulsive at shorter distances
and attractive at
larger distances, which results in the minimum of
$F_s(l)$ roughly at
$l\sim\sqrt{\lambda D}\approx d\sqrt{N}$ ($D=Nd$).
For $\alpha_s =1$ we find
$$F_s(l)\approx {{K_s n^{(1)}n^{(2)}d^2}\over{16(\pi\lambda D)^{3/2}}}
\left(2\pm\left(2-{{l^2}\over{\lambda D}}\right)\exp(-l^2/4\lambda D)
\right)\eqno(8.8)$$
This behavior of the interaction potential  can be
understood by comparing the edge profiles for large and small
surface tension.
In the former case the
profile is a monotonic function of the distance (Fig.20)
and like dislocations
repel each other minimizing the free energy associated with
the surface tension. In the case of small surface tension,
surface energy associated with curvature dominates,
the profiles become nonmonotonic functions (Fig.23)
of the distance and
like dislocations attract each other at large
separations. For $\alpha_s\ll 1$
the attraction is even more pronounced.

Finally let us discuss the interactions between two screw dislocations
in a finite system in a harmonic approximation.
The solution given by Eq.(5.14) satisfies also the boundary condition (6.5)
at the free surface. Thus we can use it for the calculation of the
distortion energy in the system bounded by two free surfaces (Eq.(6.4))
The surface induced distortion energy
per unit length
 is as follows:
$$\eqalign{F_s(l)= &\gamma (n^{(1)})^2d^2/(\pi D) \ln{(L_\rho/r_c^{(1)})}+
\gamma (n^{(2)})^2d^2/(\pi D) \ln{(L_\rho/
r_c^{(2)})}\cr \pm &2\gamma n^{(1)} n^{(2)} d^2/(\pi D)
\ln{(L_\rho/l)}\cr}.\eqno(8.9)$$
 Here $L_\rho$ is the horizontal size of the system
and $D$ is the thickness of the film.
Bulk terms (5.15),(5.16) and (8.3)
are not included in Eq(8.9). The like screw dislocations repel each other
and the unlike dislocations attract each other, contrary to the
case of bulk induced interactions (8.3). In both cases the potential is
long ranged and has the same logarithmic form.
The bulk term
(8.3)
is much larger than the surface term even for
very small $D\sim 10 d$. However it should be noted that (8.3) has been
obtained from the anharmonic terms. At the harmonic approximation
the bulk interactions are zero and then (8.9) is the only contribution to the
interaction potential.

\centerline{\bf IX. Fluctuations induced interactions between dislocations}

Dislocations are low-dimensional manifolds just like
strings$^{61}$, membranes in solutions$^{8,62}$, steps at the
solid-vacuum interfaces$^{63}$, polymers$^{64}$, interfaces$^{65}$
and domain walls in
monolayers$^{66}$. This fact has not been fully recognized so far(
see however$^{7,67,68}$).
Here we would like to explore the consequences of
the flexibility of dislocations on their behavior in finite
lamellar films. Such studies are important for
understanding the behavior of walls formed by dislocations
in liquid crystals, diblock copolymers or lamellar phases
of microemulsions, formation of meniscus in lamellar
phases, interactions and mobility in systems of dislocations.

{}From the seminal work of Helfrich$^{62}$ it follows that
flexible strongly fluctuating objects such as membranes
experience long range repulsive forces when brought
together$^{8}$.
They are called undulation forces and arise from
strong fluctuations and short range repulsion between
the objects. In the case of membranes in aqueous solution
this short range repulsion is due to the hydration forces$^{69}$.
Here we shall provide the same analysis and discuss these forces for
edge dislocations.

The single edge dislocation of length $L_y$
located along the $y$ axis and fluctuating in the $x-y$ plane (Fig.24)
in the bulk lamellar system can be described by the following
elastic energy:
$$H_0={1\over 2}\int dy E_0\left({{\partial u_x}\over{\partial
y}}\right)^2,\eqno(9.1)$$
where $u_x$ is the displacement along the $x$ direction (Fig.24),
and $E_0$ is the line tension given by Eq(5.7). Due to strong fluctuations
$<u_x^2>$ grows as the length of the dislocation
$L_y$. However these fluctuations are damped in the
presence of other dislocations. In a system of parallel
dislocations located in the $x-y$ plane and separated from one another
by a
distance $l$,
the full elastic energy:
$$H={1\over 2}\int dx dy\left ({{E_0}\over l}
\left({{\partial u_x}\over{\partial
y}}\right)^2+B_0\left({{\partial u_x}\over{\partial
x}}\right)^2\right),\eqno(9.2)$$
where the dislocation compressional modulus
$B_0=(3\pi k_BT)^2/(16 E_0 l^3)$ arises self
consistently from the fluctuation free energy, in the same way
as in a stack of membranes$^{62}$. Here $l$ is the average
separation of edge dislocations.
Following Helfrich$^{62}$ we find, using Eq(9.2),
the undulation interactions
per unit length, between the dislocations:
$$F_{H}(l)={{3\pi^2(k_BT)^2}\over{32 E_0 l^2}}\eqno(9.3)$$
Eq(9.3) differs from Helfrich result only by a numerical factor.
As we discussed above, short range repulsion must be present in
the system in order for the Helfrich mechanism to hold.
In general the
source of the short range repulsion can come from the
core
structure.
The edge
dislocation with the core splitted into two
disclinations (Fig.4c) is indeed characterized by large core
energy and complex core structure and consequently
two such dislocations of opposite or same Burgers vectors might
repel each other at short distances.

The fluctuations of the screw dislocations are more subtle
than those of edge dislocations.
In particular the line tension is not
equal to the energy of a straight screw dislocation per unit length
(compare Eq(5.19-20)), but is much larger$^{44}$.
The crucial point here is that in the process of deformation
the screw dislocation
aquires an edge character. This is the main source of the
large configurational tension given by Eq(5.20).

\centerline{\bf X. Unbinding transition for edge dislocations}

Our strongly fluctuating objects can also attract each other
(e.g. membranes interact via the van der Waals potential)
and at large distances this interaction may compete with
undulation forces. At
sufficiently low temperatures or sufficiently strong
attraction the fluctuating objects stay close minimizing
their energy. When the temperature is raised or attraction
weakened they separate, maximizing their entropy.
The transition from the bound to the free state is called
the unbinding
transition$^{70}$. As follows, there are three
elements necessary to induce unbinding transition:
strong fluctuations of objects, short range repulsive
forces and sufficiently
strong attraction between the objects at large distances.
We have already discussed them in the previous sections. Here
we will show that edge dislocations can undergo the unbinding
transition at suitable conditions$^{71}$.

Since for opposite dislocations the elastic interaction (Eq(8.7)) is
attractive, this
competes with the Helfrich repulsion given by Eq(9.3).
In the extreme limit of $\alpha_s\to\infty$
(which correspond to the system sandwiched between two
walls) we find that, as a result of this competition,
the opposite dislocations are
stabilized at the distance,
$l_{eq}=\pi((3
(k_BT)^2D)/(2d^2BE_0))^{1/3}$.
This distance corresponds to the minimum
of the sum of undulation and elastic interactions
$F_H(l)+F_s(l)$ (here for simplicity we assumed $n^{(1)}=n^{(2)}=1$).
One finds for the typical values of the smectic parameters and $D=100\mu$m
that this distance is very small, i.e. $l_{eq}=100$\AA.
It is understandable that we cannot expect an unbinding
transition in this case, since in the limit of hard walls
the elastic interations grow linearly as a function of $l$.
For finite $\alpha_s$, although in principle the
unbinding transition would be possible in thin films, we
find for typical smectic liquid crystal parameters
($\gamma =30$ dyn/cm, $\sqrt{KB} =6$ dyn/cm,
$k_BT =4\cdot 10^{-14}$ erg (room temperature),
$\lambda=d=30$\AA) that it does not occur.
For all film thicknesses the opposite dislocations
are stabilized at finite separations, providing
that the Helfrich mechanism holds in their case.
In general these results tell us that in the case of large
surface tension elastic interactions are much stronger than
the Helfrich repulsion.

For small surface tension (comparable to $\sqrt{KB}$)
the undulation interactions (Eq.(9.3)) compete with the
attractive potential for like dislocations (Eq(8.8)) in
thick films (large $D$).
As the thickness of
the film is increased the undulation interactions
win and the unbinding transition occurs.

In our calculations we have used the same parameters
as above, assuming
$\alpha_s\approx 1$, $2>K_s/(Kd)>1$ and
$n^{(1)}=n^{(2)}=1$. We have also assumed
$r_c=d$ and $E_c=\sqrt{KB} d^2/(2 r_c)$.

We have noted that practically for $\alpha_s>\alpha_s^*=1.5$ no bound
states for dislocations exist for any film thicknesses  (for $K_s/K=1$).
This critical value of $\alpha^*_s$ depends on the surface
bending elastic constant e.g. for $K_s/Kd=10$
we find $\alpha_s^*=4$.

In Fig.25
the plot of $F(l)=F_s(l)+F_H(l)$ versus $l$ is shown for
three film thicknesses ($N=$30,40 and 54) ($D=Nd$).
As the thickness
of the film is increased the value of the energy
at the minimum at $l_0\approx 3 d\sqrt{N}$
approaches zero. The unbinding transition for $K_s/Kd=1$
and $\alpha_s=1$ takes place at $N=54$, when
$F(l_0)=F(\infty)$. For thicker films
($N>54$) the dislocations are free. For thinner
films ($N<54$) they are bound.

In Fig.26 we show the diagram
of $K_s/Kd$ versus $N$. The dotted line separates the
regions where the unbound state is expected from the one
where the dislocations are bound. For large surface
bending elastic constant we find the
unbinding transitions at larger thicknesses of the film.
On the contrary for larger surface tensions we find
the unbinding transition takes place for thinner films
(Fig.27).

One notes that the transition is
extremely sensitive to the value of the surface tension
for $\gamma\approx\sqrt{KB}$.
Changing it by 1\% we get the decrease of $N$ by a factor of 2.
We can check that at the room temperature changing the
temperature by one degree upwards decreases the thickness
at which the unbinding transition takes place
(for $\alpha_s=1$,$K_s/Kd=1$) from $N=54$ to $N=52$.
Therefore there are four parameters which affect the unbinding
transition: surface tension, surface bending elastic
constant, film thickness and temperature. All of them can
be controlled, and so our results
can be tested experimentally, especially that
the number of soft condensed matter
systems characterized by small surface
tension is quite large.

\centerline{\bf XI. Unbinding transition for dislocation loops}

Dislocation loops in bulk smectic systems have been
first studied by Kl\'eman$^{72}$. Later
Helfrich$^{67}$  argued that
the nematic-smectic (NA) phase transition
in liquid crystals could be initiated by the
unbinding of dislocation loops. The mechanism would be similar to the
one proposed by Kosterlitz and Thouless (KT) for the phase transitions in two
dimensions$^{73}$.
As shown by
Nelson and Toner$^{68}$ the smectic with unbound loops
behaves like a nematic. The estimate of the transition temperature
in the loop mechanism is simple and similar to the KT estimate. The free
energy of the loop in the bulk has the following form:
$$F_{loop}=F_c L-k_BT{L\over d}\ln{p},\eqno(11.1)$$
where $L$
is the total length of the loop, $F_c$ is the
self energy of the loop per unit length
and $p$ is the number larger than one
(for the problem considered
on the lattice this number
would be close to the coordination number of the lattice).
Here $k_B(\ln p)/d$ is the entropy of the loop per unit length;
we count the number of loop configurations and neglect
the entropy of placing
and closing the loop. The latter is proportional
to the log of the size of the loop  and in general controls only
the density of loops but does not influence the transition
temperature,
$$T_{NA}=F_cd/(k_B\ln{p}).\eqno(11.2)$$
At this temperature $F_{loop}$ changes sign
and the spontaneous growth of loops
occurs. Then, according to Nelson and Toner, the
layered smectic structure is destroyed.
A very instructive approach to this problem can be found in Ref.7.
Here we shall estimate the influence of the boundaries on the
loops, their growth and size.

Let us consider a smectic liquid crystal film of thickness $D$, bounded by two
surfaces located at $z=\pm D/2$, with a single circular dislocation
loop of radius $R$ located in the middle of the film at $z=0$.
The loop is described by the following condition
for the vertical displacement , $u({\bf r}_\bot,z)$,
of smectic layers:
$$u({\bf r}_\bot,z=0)=\cases{0, &if $\vert{\bf r}_\bot\vert
\ge R$;\cr {\rm sgn}(z)d/2,&if
$\vert{\bf r}_\bot\vert<R$.\cr}\eqno(11.3)$$
Since in the smectic liquid crystal the surface tension is larger
than $\sqrt{KB}$ ($\alpha_s>1$)
the dislocation loop is stabilized in the middle of
the film.
The solution of Eq(5.3) consistent with the condition (11.3)
and satisfying the boundary condition (6.5)
(and similar condition for the second surface obtained from (6.5)
by changing the sign of z).
is given by $u_{eq}({\bf r}_\bot,z)$,
$$u_{eq}({\bf r}_\bot,z)=u_b({\bf r}_\bot,z)+u_p({\bf r}_\bot,z)\eqno(11.4)$$
where$^{72}$
$$u_b({\bf r}_\bot,z)={\rm sgn}{(z)}{{dR}\over{4\pi}}
\int d{\bf q}_\bot
\exp{(-\lambda q^2\vert z\vert)}{{\exp{(i{\bf q}_\bot{\bf r}_\bot)}}
\over{q}}{\rm J}_1(qR)
\eqno(11.5)$$
and$^{74}$
$$u_p({\bf r}_\bot,z)={{dR}\over{4\pi}}\int d{\bf q}_\bot
{{\exp{(i{\bf q}_\bot{\bf r}_\bot)}}
\over{q}}{\rm J}_1(qR)
\left(\exp{(\lambda q^2 z)}-\exp{(-\lambda q^2 z)}\right) f(q).
\eqno(11.6)$$
Here $q=\vert{\bf q}_\bot\vert$ and $f(q)$ is defined after
Eq(8.4).
Inserting our solution
into the distortion energy $F_b+F_s$ (Eq(5.4,6.4))
we find the distortion energy of the
smectic film of thickness $D$ with the circular dislocation
loop of radius $R$ in the middle of the film:
$$\eqalign{F(D,R)=&2\pi RE_c+\sqrt{KB}d^2{\pi\over 2}R^2\int_0^{2\pi/r_c}dq q
{\rm J}_1^2(qR)\cr-&2\sqrt{KB}\pi d^2R^2\int_0^{\infty}dq q {\rm J}_1^2(qR)
f(q)\cr}.\eqno(11.7)$$
In the limit of $R\gg d$ we find the following form of the
distortion energy $F_c(D)$ per unit length as a function of the film
thickness, D:
$$F_c(D)=\lim_{R\to\infty}{{F(D,R)}\over{2\pi R}}
=E_c+{{\sqrt{KB}d^2}\over{2r_c}}
-\sqrt{KB} d^2\int_0^{\infty} dq f(q).\eqno(11.8)$$
The first two terms of the energy are associated with the bulk distortions
(Eq(5.7)) and
do not depend on the surface tension or the film thickness.
In fact they are the same as for the edge dislocation.
Thus the local distortions induced by a large dislocation
loop are the same as the ones induced by an
edge dislocation.
The sum of $E_c$ and $\sqrt{KB}d^2/2r_c$ is equal to $F_c$ (Eqs(11.1,2)).
The last term in Eq(11.8)
is associated with the finite size of the system and
the surface tension. In the limit of $D\to\infty$ it vanishes and consequently
$F_c(D)$ approaches $F_c$ as it should.
In order to estimate the NA transition temperature
as a function of the film thickness we use
$F_c(D)$ in Eqs(11.1,2).
In general, $p$ should also change with $D$. For thin films this number
 should be
close to 4 (coordination number for the two dimensional square lattice)
while for thick ones it  will be close to 6.
Here we neglect this contribution
to Eq(11.1). From Eqs(11.1,11.8) we find that$^{48,74}$:
$${{T_{NA}(D)-T_{NA}(\infty)}\over{T_{NA}(\infty)}}={{\sqrt{KB} d^2}\over
{\sqrt{\lambda D} F_c}} C(\alpha_s,K_s/(KD)).\eqno(11.9)$$
where
$$C(\alpha_s,\eta)=-\int_{0}^{\infty} dq {{(1-\alpha_s-\eta q^2)}\over
{(1+\alpha_s+\eta q^2)e^{q^2}+(1-\alpha_s-\eta q^2)}}\eqno(11.10)$$
and $T_{NA}(\infty)$ is given by Eq(11.2).
The function $C(\alpha_s,K_s/KD)$, depends on $K_s/(KD)$
very weakly, thus the
dominant dependence of $T_{NA}(D)$ on $D$
is $1/\sqrt{D}$. For $\alpha_s=5$ (typical
for smectic liquid crystal$^{51}$) we find C=1.25.
For the purpose of rough estimate
we neglect the core energy $E_c$ and set $C=1$ in
Eq(11.9), finding:
$${{T_{NA}(D)-T_{NA}(\infty)}\over{T_{NA}(\infty)}}=
{{2r_c}\over{\sqrt{\lambda D}}}.\eqno(11.11)$$
The transition temperature is larger in thin films than in the bulk,
thus the
smectic phase is stabilized in films and near free surfaces.
The large surface tension
and
finite size effects are responsible for this behavior.
For three layer film, $D=3d$, $\lambda\approx d=30$\AA, and
the core radius $r_c$ equal to the width of the molecule,
$r_c\approx$3\AA (chosing the smallest length scale in the problem
is consistent with $E_c=0$; usually it is assumed that $r_c\approx d$;
however we may encounter situations when the size of the core
can be smaller$^{39}$ or larger (see Fig.4b)),
the
transition temperature
can be shifted by tens of degrees upwards in comparison to the bulk.
It means that it should be relatively easy
to overheat a smectic film.
It must be realized however, that for large temperature shifts it is not
justified to neglect the temperature dependence of the
smectic elastic constants.

Growth of the dislocation loops is only one particular mechanism of
the nematic smectic phase transition. However it can be preempted by
some other mechanisms$^7$.
In thin films different mechanisms can be characterized
by different temperature dependence on $D$.
For the loop unbinding mechanism we have already obtained
$\delta T_{NA}=(T_{NA}(D)-T_{NA}(\infty ))/
T_{NA}(\infty)\sim 1/\sqrt{D}$. For the
3D XY universality class$^{75}$
$\delta T_{NA}\sim D^{-1/\nu}$
with$^{76}$ $\nu\sim 0.7$, by comparing the correlation length to the
thickness of the film. Finally for the first order
phase transition we find$^{77}$
$\delta T_{NA}\sim 1/D$.
Since for each model of NA transition we find
different behavior of $T_{NA}(D)$, this temperature
can serve as yet another test for the order and mechanism of NA transition.

So far we have tacitly assumed that $\gamma$ is finite.
If we set $\gamma\to\infty$, the surfaces become rigid and
we obtain the film sandwiched between solid boundaries.
The distortion energy of the loop located between
solid boundaries is proportional to the area of the loop, contrary to the
previous case of the freely suspended film when the loop distortion energy
has been
proportional to  the length of the loop.
Since $F(D,R)$ grows now
as $R^2$ while the entropy still as $R$, we conclude
(see Eq(11.1)) that the loop mechanism is supressed in this case for
any finite $D$.
Now let us estimate the size
of the loop in the film contained between solid boundaries. At
$T>T_{NA}(\infty )$ the loop can grow up to the point
when $F_{loop}$ changes sign from negative to positive.
This happens for the finite radius of the loop, $R_{eq}$.
For $D,R\gg d$ and $\gamma\to\infty$ we can make the following approximations:
$$-\int_0^{\infty}dq q {\rm J}_1^2(qR)
f(q)\approx {A\over{\lambda D}},\eqno(11.12)$$
and
$$\int_0^{2\pi/r_c}dq q {\rm J}_1^2(qR)\approx {2\over{R r_c}}.\eqno(11.13)$$
Here $A\approx 0.5$ is a constant.
Now using Eq(11.1) and Eqs(11.7,8) we find that $F_{loop}=0$
at $R=0$ and at
$$R_{eq}={{\lambda D F_c}\over{A\sqrt{KB} d^2}}
{{T-T_{NA}(\infty)}\over{T_{NA}(\infty)}}\eqno(11.14)$$
As we see for $D\to\infty$ we get $R_{eq}\to\infty$.
Certainly before $R_{eq}$ could grow to infinity
the smectic phase would undergo a transition to the nematic phase.
The mechanism of this phase transition would be
different from the loop unbinding mechanism.
However in the overheated smectic between solid boundaries one
should observe the loops of size given by Eq(11.14).

Our approach to the loop deformations neglected completely
the fact that the loop can have an edge as well as screw parts.
Certainly for large thicknesses of the film the screw character of the
loops should be visible$^{78}$. In the experiments conducted on
lamellar phases of non-ionic surfactants the density of loops increased
at the approach to the isotropic-lamellar phase transition$^{78}$.
In fact large number of dislocation loops and their interactions
might modify the transition temperature. This
is not taken into account in our approach.

\centerline{\bf XII. Dynamical properties of dislocations: mobility}

Dislocations can move inside a sample under the action of an external stress
or stress created by {\bf another} dislocation. Let ${\bf\sigma}^E$ be this
stress (excluding the stress created by the dislocation itself). The
force ${\bf F}^E$ acting on the dislocation (per unit length)
can be calculated
similarly as in solids from the Peach Koehler formula$^{33}$:
$${\bf F}^E_i=\epsilon_{ikl}{\sigma}^E_{lm}b_mt_k\eqno(12.1)$$
where ${\bf b}$ is the Burgers vector of
the dislocation (of length $\vert{\bf b}\vert=nd$)
and ${\bf t}$ is the unit vector along the dislocation
line. The stress tensor is given by Eqs.(4.6-8) (see also Ref[43]).
This force sets the dislocation in motion. If the dislocation
moves slowly with respect to the sound velocity the effective inertial
forces can be neglected, since the effective mass is very small in this case
(typically a row of molecules). Thus the equation of motion can be
simplified to the following form:
$${\bf F}^E+{\bf F}^V=0,\eqno(12.2)$$
where ${\bf F}^V$ is the viscous force (section IV). Now based on Eq(12.2)
we can discuss the motion of screw and edge dislocations. Since  the
distiction between these two types of dislocations are more pronouced in
smectics than in solids we shall consider their mobility separately.
Let us first consider the edge dislocations. There are two basic
kinds of motion: the glide and climb. The former is
perpendicular, while the latter is parallel to the layers.
These two kinds of motion are different: glide is conservative, while climb
requires transport of matter by diffusion (permeation). In solids
glide is favored and in fact is responsible for plastic deformations
at low temperatures ($T<T_m/2$; $T_m$ is the melting temperature).
By contrast climb seems to be  favored in smectics (at least
in thermotropic smectics) and can be induced by the compression or dilation
of layers while glide requires bending of the layers.
Let us first consider the climb motion of an edge dislocation in a smectic
system
submitted to  the normal stress $\sigma_{zz}^E=\sigma$ (Fig.28).
The force acting on the dislocation of Burgers vector length
$\vert{\bf b}\vert=nd$ is (Eq(12.1)):
$$F^E=-\sigma nd,\eqno(12.3)$$
while the friction force is
$$F^V=-{{nd v}\over{m}}.\eqno(12.4)$$
{}From Eq(12.2) we find the velocity:
$$v=-m\sigma,\eqno(12.5)$$
where $m$ is the mobility of the dislocation, a  priori being a function of
the velocity, the length of the Burgers vector, $nd$, and the material
constants. The mobility of an elementary dislocation$^{79}$
is related to the
self diffusion coefficient, $D_\parallel$,
for thermotropic smectics
assuming that the mean distance between jogs, $l$,
along the line is comparable to
the molecular size. This assumption is reasonable in thermotropic smectics
and gives  with analogy to solids  (metallurgical model):
$$m\approx{{D_\parallel v_{\rm mol}}\over{k_BT l}}.\eqno(12.6)$$
Here $v_{\rm mol}$ is the molecular volume and $l\sim d$.
This model is not correct for giant dislocations, especially those
with core splitted into two $\pm 1/2$ disclinations.
In this case it is better to assume that the dislocation acts as an obstacle
of width $nd$ perpendicular to layers (Fig.29).
The flow of a smectic around such a ribbon leads to four permeation
boundary layers in which dissipation occurs due to permeation$^{80,81}$.
A calculation of dissipation gives the force of friction acting on the ribbon
and the following expression for the mobility of the dislocation:
$$m=\sqrt{{\lambda_p}\over{\mu}},\eqno(12.7)$$
where $\lambda_p$ is the permeation coefficient and $\mu$ is the shear
viscosity (see section IV).
A more sophisticated calculation$^{82}$ taking into account
deformations of the layers and assuming strong permeation
($l_p=\sqrt{\lambda_p\mu}>\lambda=\sqrt{K/B}$)
gives the following formula:
$$m=A{{\lambda}\over{d}}\sqrt{{{\lambda_p}\over{\mu}}},\eqno(12.8)$$
where $A$ is a numeric  factor of the order of one.
The mobility does not depend on
the Burgers vector nor on the velocity, due to the fluidity of the layers.
The two formulas (12.6) and (12.8)
are compatible since $\lambda_p\sim D_\parallel v_{\rm mol}/k_B T$
(section IV)
and $\mu\sim k_BT/l D_\parallel$.

Let us now describe the glide motion of an edge dislocation.
In this motion, the dislocation moves perpendicularly to the layers (Fig.30).
Two cases must be considered. Below the Peierls-Nabarro stress
the dislocations remain pinned in its Peierls valley (at zero temperature).
For smectics this stress has been calculated by Lejcek $^{83}$:
$$\sigma_{PN}={3\over 4} B\sqrt{{{2\pi\lambda}\over{r_c}}}\exp{(-\pi r_c/nd)}.
\eqno(12.9)$$
The process of glide must be thermally activated for
$\sigma^E_{zz}<\sigma_{PN}$. It requires the formation of small bulges
(Fig.31) which can afterwards spread sideways.
For stresses larger than $\sigma_{PN}$ dislocations can glide easily
and the mobility is$^{84}$
$$m={{16r_c\sqrt{\pi\lambda r_c}}\over{\mu nd(1+3\lambda/(2 r_c))}}.
\eqno(12.10)$$
The glide mobility is of the same order of magnitude as the climb mobility
Eq.(12.8).
The more dissociated the core is ($r_c$ large, Fig.4b) the easier
glide proceeds (large mobility). However for the giant dislocations
with core splitted into $\pm 1/2$ disclinations the glide is much more
difficult than climb, since the core is extended along the $x$ direction.

To finish this section let us give the mobility of screw dislocations.
These  defects can easily glide  without the layer resistance. The mobility is
given by the following formula:
$$m={{8\pi^2r_c^2}\over{\mu nd}}.\eqno(12.11)$$
The motion of the screw dislocation is conservative and thus permeation is
negligible. The friction force is due to the viscosity of the medium
$\mu$.

\centerline{\bf XIII. Microplasticity and helical instability
of screw dislocations}

Experimentally  we can reveal the dynamical properties of dislocations by
compressing a homeotropic wedge sample (see section II)
and recording its viscoelastic response.
More precisely one can impose a step like variation or a
sinusoidal variation
of the thickness
with piezoelectric ceramics and
measure the normal stress as a function of time. The first experiment
of this type
was performed by Bartolino and Durand$^{85}$ in smectic-A liquid crystal.
By applying very small thickness variation (a few {\AA} or less) they
showed for the first time the elastic behavior of smectics in compression
normal
to layers and its plastic relaxation in time.
Ribotta also studied with this system the undulation
instabilities of layers in dilation and their subsequent
instabilities leading to focal parabolas$^{34,86}$. In order to
study dislocationsand plasticity  a new cell with stackings of
piezoelectric ceramics, allowing for much larger
deformations, was used$^{87}$.
The step like thickness variations (ranging from $d$ to $10d$ typically)
and the normal stress as a function of time are shown schematically
in Fig.32ab.
The stress first jumps abruptly to
$\sigma_0\approx B\delta_0/d$ (elastic regime) and then relaxes exponentially
to zero with the characteristic time $\tau$. This time depends on the
initial stress$^{88}$ (Fig.32c) and decreases by
successive jumps with the increase of
$\sigma_0$.
Each jump is associated with the change of the thickness
by $N_p=p N_1$ layers ($p$ is the index of the jump)
and by the relaxation time $\tau_p$
defined by:
$${1\over{\tau}}={1\over{\tau_g}}+{1\over{\tau_p}}.\eqno(13.1)$$
The time $\tau_g$ is the relaxation time at small deformations.
The experiment shows that $N_1\sim 3$ independent of the thickness of
the sample
$D$ and the angle $\alpha$ between the plates. One finds from the
experimental results that $\tau_g\sim D/\alpha$ and
$\tau_p\sim D/p$, where $p$ is the index of the jump (Fig.32c).
Also $\tau_p$ is independent of $\alpha$.
Two relaxation mechanisms are involved in the observed phenomena.
The first one is the motion of edge
dislocations which are present in the sample
because of the wedge geometry. The relaxation time for this motion is
$$\tau_g={D\over{mB\alpha}}\eqno(13.2)$$
proportional to $D/\alpha$ in agreement with experiment. From (13.2) we
can extract the mobility. For 8CB at room temperature
one finds $m\sim 5\cdot 10^{-7}$cm$^2$s/g
and from Eq.(12.7) we
calculate $\lambda_p\approx 5\cdot 10^{-13}$ cm$^2$/poise
($\mu\approx 2$poise). This gives $l_p=\sqrt{\mu\lambda_p}\approx 10${\AA}.
This length is of the order of molecular size as expected from Eq.(12.6)
and Einstein relation for viscosity.

The second mechanism which explains the jumps of the relaxation
time is the sudden instability of screw dislocations under the applied
deformation$^{44}$.If
$$\sigma_0>p {{2\pi T_{\rm screw}}\over {D}}\eqno(13.3)$$
then, each elementary
screw dislocation ($n=1$)
joining two plates (and anchored to them)
is unstable with respect to the formation of a helix of pitch $D/p$.
This helical shape is equivalent to the removal (under compression)
of $p$ layers inside the cylinder in which the screw dislocation is located.
(Fig.33). That is why the relaxation time is inversely proportional to
the jump order $p$. Finally we can use the formula for $N_1$:
$$N_1=2\pi {{T_{\rm screw}}\over{Bd^2}}\eqno(13.4)$$
to estimate $E_c$ using the theoretical value of $T_{\rm screw}$ (5.19-20).
One finds $E_c\simeq 0.1 B d^2$. This value of the core energy is
compatible with the one obtained for  the core filled with nematic.
The relaxation time
associated to jump p can also be calculated:
$$\tau_p\simeq {{d^2 D}\over{2\pi^2 m T_{\rm screw}p}}\eqno(13.5)$$
where $m$ is the screw dislocation mobility given by Eq.(12.11).
{}From Eq(13.5) and $\tau_p$ measurements
we find $m\sim 10^{-6}$cm$^2$s/g in agreement with
Eq(12.11).

So far we have assumed that both relaxation processes are independent,
but we know that edge dislocation must cross the screw dislocation in the
process. This crossing leads to the formation of the jogs
on the edge dislocation (Fig.34).
The jog has the screw character so its energy
(Eq(5.15)) is
$$E_{\rm jog}\sim {{B d^3}\over{64\pi^3}}\eqno(13.6)$$
assuming $n=1$ for both dislocations and $r_c=\lambda=d$. Experimentally
$B\sim 5\cdot 10^7$dyn/cm$^2$ and $d\sim 30${\AA}, thus $E_{\rm jog}\sim
10^{-2}
k_BT$. This jog can be very easily thermally activated thus
elementary dislocations can very easily cross. Also two
edge dislocations interact only when they are in the same slip plane which
is rarely the case, consequently each dislocation moves independently from
the other dislocations, contrary to the case of dislocations in solids.
Of course once again fluidity of the layers is responsible for
this behavior of dislocations. Finally, one may ask why helical instability
of screw dislocation is more favorable than the nucleation ex-nihilo
of dislocation loops. The answer is as follows. The length increase
of the screw dislocation when it aquires a helical shape of radius $R$
is $2\pi R\times \pi R/D$, whereas the nucleation of the
dislocation loop of equivalent radius requires to make a length
$2\pi R$  of edge dislocation. The helical instability is thus much more
favorable energetically than the nucleation ex-nihilo of dislocation
loop.

Dislocations may also play an important role in the process of growth.
This has been observed at the smectic A-smectic B interface.
When the smectic B grows from the undercooled smectic A, it generates stresses
that are mainly relaxed by the motion of dislocations in the
smectic A phase.
This phenomenon is detectable owing to the entrapment of dust particles
which are dragged away when they climb parallel to the layers.
If the growth rate is too large, then dislocations may no longer
relax the stresses efficiently and then smectic A breaks with the formation
of many focal conic domains$^{84}$.

So far we have considered measurements performed on smectic liquid crystals.
Similar measurements have been done for self assembling
amphiphilic systems of C$_{12}$E$_5$ (pentaethylene glycoldodecyl ether)
and water. Surprisingly this lamellar system orients very
easily and spontaneously
in homeotropic anchoring. For example few minutes are enough to obtain a good
homeotropic
sample without visible defects, whereas several weeks of annealing are
necessary
to obtain similar results with phospholipid lamellar phase$^{26}$.
This simple observation shows that permeation is much faster in
C$_{12}$E$_5$ than in usual lyotropic systems, the fact also confirmed by the
direct measurements of the edge dislocation mobility$^{89}$.
The mobility has been found of the same order of magnitude as in thermotropic
liquid crystals considered previously in this section.
The screw dislocations are very numerous in this lyotropic
system$^{90}$ and presumably they favor permeation. In fact it was shown that
these defects act as vortices when a pressure gradient is imposed
perpendicularly
to the layers. This mechanism leads to the effective permeation
coefficient$^{91}$
$$\lambda_p={{A(nd)^4}\over{\mu L^2}}\eqno(13.7)$$
where $L$ is the mean distance between dislocations and
$$A=0.07+0.009\ln{L/r_c}.\eqno(13.8)$$

Finally we note that dislocations strongly affect flows as will be
shown  in the next section.

\centerline{{\bf XIV. Lubrication}$^{92}$}

It is well known that lamellar phases are very good lubricants. As an
example one can
cite soaps, graphite,
and also lamination oils, which under high pressure and temperature
acquire a lamellar structure.  In the classical theory, lubrication
occurs when a fluid is sheared between two planar surfaces making a
small angle
$\alpha$. The lubrication force is due to the increase of pressure in
the locally compressed fluid. By replacing the isotropic fluid with smectic A
one
expects similar,
but more spectacular effects, since apart from the easy flow of the layers
past
each other, there is a genuine solid like strength perpendicular to layers.
Fig.35 shows the lubrication geometry. The sample is sandwiched between
two plates with homeotropic anchoring at an angle $\alpha$. In this sample
as we know there is an array of dislocations at a mean distance $d/\alpha$.
At time $t=0$ the upper slide starts to move with the horizontal velocity $v$.
At time $t$ the thickness of the sample  has varied perpendicularly to a fixed
position by an amount $\Delta D=-vt\alpha$, which is positive (dilation)
when $v<0$ and negative (compression) when $v>0$. This thickness variation
creates a normal stress $\sigma_{zz}^E$ which makes the edge dislocations
to climb together
in a direction which favors the relaxation of the imposed deformation.
The dislocations are dragged by the flow with velocity $v/2$ at time $t=0$
and then progressively slow down until they stop in the stationary regime
($t\to\infty$).
A direct calculation of the normal stress gives:
$$\sigma_{zz}^E=-{{v}\over{2m}}\left(1-\exp{(-t/\tau_g)}\right),\eqno(14.1)$$
with $\tau_g$ given by (13.2).
The  stress exerts a normal force on the slides, which adds to the
one exerted by the hydrostatic pressure. Both terms can be calculated
in the limit of a very small angle $\alpha$.
In the limit $t\gg\tau_g$ one finds$^{92}$ the force per unit length
$$F_n=F_{\rm hydrostatic}+F_{\rm smectic}=
{1\over 2}\mu\left({L\over D}\right)^3 v\alpha+{{Lv}\over {2m}}.
\eqno(14.2)$$
Here $L$ is the horizontal size of the sample.
The first term is related to the hydrostatic
pressure and is always present, even for isotropic lubricants, whereas
the second one is specific for layered structures. In fact the second term
can be much larger than the first one. We estimate
$F_{\rm hydrostatic}=10^3v$[dyn s/cm$^2$] and
$F_{\rm smectic}=10^6v$[dyn s/cm$^2$], assuming
$L=2$cm, $D=200\mu$m, $\mu=1$poise, $\alpha=2\cdot 10^{-4}$ and
$m=10^{-6}$cm$^2$ s/g. Since the normal force is three orders of magnitude
larger in the case of smectics it means that
smectics better keep the two moving surfaces apart and consequently are
better
lubricants than an ordinary fluid of the same viscosity.
Of course the full comparison of the lubrication properties involves the
calculation of the force  of friction. Assuming simple flow
pattern $v_x=v z/D$ we find the  force of friction (per unit length)
$$F_{\rm friction}=-\mu {{v L}\over D}-
{{\alpha L v}\over{4m}},\eqno(14.3)$$
and therefore the smectic behaves as a fluid with
apparent viscosity
$$\mu_{\rm app}=\mu+{{\alpha D}\over{4m}}\simeq \mu+{{\alpha D\mu}
\over{4l_p}}.
\eqno(14.4)$$
Thus the corrections to viscosity $\mu$ can be very significant even
for very small angles $\alpha$,
because $D$ is much larger then the permeation length
$l_p\sim d$ (see section XII before Eq.(12.8)). For  instance $\mu_{\rm app}=
5\mu$, when $\alpha=2\cdot 10^{-4}$rad, $D=200\mu$m and $l_p=25$\AA.
This larger viscosity is of course not
favorable for lubrication.
Another consequence of this calculation is that lubrication
effects must be taken into account when one measures the shear viscosity
with rheometer. Nevertheless, it is possible, at certain conditions,
to completely eliminate the lubrication effects and to measure the
intrinsic shear viscosity of the sample. Even in this limit, the
viscosity appears to be strongly dependent on the shear rate $S=v/D$;
large at small $S$, $\mu_{app}$ approaches $\mu$ at large $S$ (Fig.36).
Careful
observations of the motion of dust particles
in the samples show that the viscosity increase at smaller shear rates
is related to strong departure from the linear velocity profile
(Fig.37). In order to explain the rigidification of the velocity profile
in the
middle of the sample a mechanism of strong interaction between screw
dislocation and the
flow has been proposed$^{93}$. Their stationary shape can be obtained by
comparing
the viscous torque that the flow exerts on the line with the elastic
torque due to the line tension. The
dislocations bend while
acquiring a mixed character and tend to occupy a finite fraction
of the whole thickness
of the sample, when the shear rate becomes larger than the critical
shear
$$S_c={{16B d^2}\over{\mu D^2}}\approx {{16K}\over{\mu D^2}}.\eqno(14.5)$$
The larger the shear rate the thinner is the zone where the flow is rigidified
(small velocity) (Fig.37). At very large $S$ a dislocation is completely
stretched and do not longer contribute  to the total dissipation. In this
limit the apparent viscosity $\mu_{\rm app}$  is equal to the intrinsic
viscosity
$\mu$. This model allowed us to explain most of the experimental results and
to estimate
the density of screw dislocations ($10^6$cm$^{-2}$)
in usual sample of thermotropic
smectic liquid crystals$^{94}$.

\centerline{\bf Summary}

In this review paper we have tried to summarize the progress in the theory of
dislocations which took place in  the recent decade.
Both equilibrium and nonequilibrium,
bulk and surface properties have been discussed. It has been shown that
theoretical and experimental studies are very closely related in this field
and that many interesting aspects of the behavior of dislocations
can be studied experimentally.
Many properties of dislocations in smectics are very much different from those
of solids. We have tried to emphasize all the differences.
In general we can say that
smectics have certain solid like and liquid like properties
combined in the very unsual way and in this review we have
shown how the solid-like  and liquid-like
behavior of smectics influences dislocations.

\centerline{\bf Acknowledgements}

We are grateful to CNRS for making our collaboration possible.
This work has been also partially supported by KBN under grants
2 P 30219004 and 2 P30302007. RH. acknowledges the hospitality
of Ecole Normale Superieure.

\centerline{\bf References}

\item{1.} J.Weertman and J.R.Weertman, {\it Elementary Dislocation Theory},
(Oxford University Press, New York, Oxford, 1992).
\item{2.} D.Hull, {\it Introduction to Dislocations}, (Pergamon Press second
ed.
1975).
\item{3.} see e.g. S.Amelinckx, Acta Metall {\bf 6}, 34 (1958).
\item{4.} see e.g. S.Amelinckx and W.Dekeyser, Solid State Physics {\bf 8}, 327
(1959).
\item{5.} G.Gompper and M.Schick, {\it Self-Assembling Amphiphilic Systems}
in Phase \hfill\break Transitions and Critical Phenomena eds. C.Domb and J.L.
Lebowitz
vol. 16 (Academic Press, 1994).
\item{6.} F.S.Bates  and G.H.Fredrickson, Annu.Rev.Phys.Chem. {\bf 41}, 525
(1990).
\item{7.} P.G. de Gennes and J.Prost, {\it Physics of Liquid Crystals},
(Clarendon Press, Oxford 1993).
\item{8.} D.Roux and C.R.Safinya, J.Phys. (France), {\bf 49}, 307 (1988).
\item{9.} Y.Bouligand, in {\it Physics of Defects}, Les Houches,
ed. Balian et al. (North Holland Publishing, 1981).
\item{10.} S.Chandrasekhar and G.S.Ranganath, Adv.Phys. {\bf 35}, 507 (1986).
\item{11.} M. Kl\'eman, Rep.Prog.Phys., {\bf 52}, 555 (1989).
\item{12.} P.G. de Gennes, C.R. Acad.Sci. Paris {\bf 275B}, 939 (1972).
\item{13.} R.B.Meyer, B.Stebler and S.T.Lagerwall, Phys.Rev.Lett. {\bf 41},
1393 (1978).
\item{14.} M. Kl\'eman, C.Colliex and M.Veyssie in {\it Lyotropic Liquid
Crystals},
ed S.Freiberg p. 71 (1976).
\item{15.} S.A. Asher and P.S.Pershan, Biophys. J. {\bf 27}, 393 (1979).
\item{16.} M.Kl\'eman, C.E.Williams, M.J. Costello and T. Gulik Krzywicki,
Philos.
Mag.
{\bf 35}, 33 (1977).
\item{17.} C.E. Williams and M.Kl\'eman, J.Phys.Lett. (France) {\bf 35}, L33
(1974).
\item{18.} R.Ribotta, J.Phys. Coll. (France) {\bf 37}, C3 149 (1976).
\item{19.} P.S.Pershan, J.Appl.Phys. {\bf 45}, 1590, (1974).
\item{20.} L.Lejcek and P.Oswald, J.Phys.II {\bf 1}, 931 (1991).
\item{21.} M.Maaloum, D.Ausserre, D.Chateney, G.Coulon and Y. Gallot,
Phys.Rev.Lett {\bf 68}, 1575 (1992).
\item{22.} M.S.Turner, M.Maaloum, D.Ausserr\'e, J-F. Joanny and M. Kunz,
J.Phys. II (France), 689 (1994).
\item{23.} P.Piera\'nski et al. Physica A {\bf 194}, 364 (1993).
\item{24.} F. Grandjean, Bull.Soc.Fran\c c.Min\'er {\bf 39}, 164 (1916);
C.R.Acad.Sci.Paris {\bf 166}, 165 (1918).
\item{25.} C.Williams, Th\`ese d'Etat, Universit\'e d'Orsay n. A012720 (1976).
\item{26.} W.K.Chan and W.W.Webb, J.Phys. (France), {\bf 42}, 1007 (1981).
\item{27.} G.Friedel, Annales de Physique {\bf 18}, 273 (1922).
\item{28.} J. Bechhoefer and P. Oswald Europhys.Lett. {\bf 15}, 521 (1991).
\item{29.} J. Bechhoefer, L.Lejcek and P.Oswald, J.Phys. II (France) {\bf 2},
27
(1992).
\item{30.} L.Lejcek, J.Bechhoefer, P.Oswald, J.Phys. II (France) {\bf 2}, 1511
(1992).
\item{31.} M. Allain, J.Phys. (France) {\bf 48}, 225 (1985).
\item{32.} J.A.N. Zasadzinsky, Biophys.J. {\bf 49}, 1119 (1986).
\item{33.} L.~D.~Landau and E.~M.~Lifshitz, ``
Theory of Elasticity"
(Pergamon Press, Third edition,
 1986), p. 172-177.
\item{34.} R.Ribotta, G.Durand, J.Phys. (Paris) {\bf 38}, 179 (1977).
\item{35.} R.Ho\l yst and A.Poniewierski, J.Phys. (France) II, {\bf 3},
177 (1993).
\item{36.} P.G.~de~Gennes, J.Phys(Paris) Colloq.{\bf 30},C4 65 (1969).
\item{37.} G.~Grinstein and R.A.~Pelcovits Phys.RevA, {\bf 26}, 915 (1982).
\item{38.} N.Schopohl and T.J.Sluckin, Phys.Rev.Lett.{\bf 59}, 2582 (1987).
\item{39.} S.D.Hudson and R.G.Larson, Phys.Rev.Lett. {\bf 70}, 2916 (1993).
\item{40.} S.R.Renn and T.C.Lubensky, Phys.Rev.A {\bf 38}, 2132 (1988);
\hfill\break S.R.Renn, Phys.Rev. A {\bf 45}, 953 (1992).
\item{41.} J.W. Goodby, M.A. Waugh, S.M. Stein, E.Chin, R.Pindak and J.S.
Patel,
 Nature {\bf 337}, 449 (1989).
\item{42.} S.Kralj and T.J. Sluckin, Phys Rev E {\bf 48}, 3244 (1993).
\item{43.} M.Kl\'eman {\it Point, Lines and Walls}, (Wiley 1983).
\item{44.} L.Bourdon, M.Kl\'eman, L.Lejcek and D.Taupin, J.Phys. (Paris) {\bf
42},
261 (1981).
\item{45.} M.Kl\'eman and L.Lejcek, Phil.Mag. A {\bf 42}, 671 (1980).
\item{46.} H. Pleiner, Liq.Cryst. {\bf 1}, 197 (1986); Phil.Mag. A{\bf 54}, 421
(1986);
Liq.Cryst. {\bf 3}, 249 (1988).
\item{47.} L.Lejcek, Liq.Cryst. {\bf 1}, 473 (1986).
\item{48.} R.Ho\l yst, Phys.Rev.Lett. {\bf 72}, 4097 (1994).
\item{49.} R.Ho\l yst, Macromol. Theory Simul. {\bf 3}, 817 (1994).
\item{50.} A.Poniewierski and R.Ho\l yst, Phys.Rev. B {\bf 47}, 9840 (1993).
\item{51.} R.Ho\l yst, D.J.Tweet and L.Sorensen, Phys.Rev.Lett.
{\bf 65}, 2153 (1990);\hfill\break R.Ho\l yst, Phys.Rev A {\bf 44}, 3692
(1991).
\item{52.} A.N.Shalaginov and V.P.Romanov, Phys.Rev E {\bf 48}, 1073 (1993)
{\it actually in this reference the authors miss the surface curvature term}.
\item{53.} D.J.Tweet, R.Ho\l yst, B.D. Swanson, H.Stragier and L.B.Sorensen,
\hfill\break Phys.Rev.Lett. {\bf 65}, 2157 (1990).
\item{54.} R.Bartolino and G.Durand, Mol.Cryst.Liq.Cryst. {\bf 40}, 117 (1977).
\item{55.}F.Nallet and J.Prost, Europhys.Lett. {\bf 4}, 307 (1987).
\item{56.} C.Guilliet, P.Fabre and M.Veyssi\'e,
J.Physique II {\bf 3}, 1371 (1993).
\item{57.} S.Wu, J.Phys.Chem. {\bf 74}, 632 (1970).
\item{58.} G.Henn, H.Poths and M.Stamm, Polymers for Advanced  Technology,
{\bf 5}, 582 (1994).
\item{59.} G.Henn, M.R\"ucker, J.P. Rabe and M.Stamm (in press, 1995).
\item{60.} Z.Cai, K.Huang, P.A.Montano, T.P.Russell, J.M.Bai
\hfill\break and G.W.Zajac,
J.Chem.Phys. {\bf 98}, 2376 (1993).
\item{61.} N.G. van Kampen, {\it Stochastic Processes in
Physics and Chemistry}, (North Holland Publishing Company,
1981) p. 67; R.R.Netz and R.Lipowsky, J.Phys I {\bf 4}, 47
(1994).
\item{62.} W.Helfrich Z.Naturforsch. {\bf 33a}, 305 (1978);
also in {\it Liquids at interfaces}, eds J.Charvolin,
J.F.Joanny and J.Zinn-Justin (Elsvier Science Publisher
1990), p.212.
\item{63.} P.Nozi\`eres in {\it Solids far from
Equilibrium}, ed C.Godr\`eche, (Cambridge University Press,
1992, p.1.
\item{64.} P.G.de Gennes {\it Scaling concept in polymer
physics} (Cornell University Press, 1979).
\item{65.} G.Forgacs, R.Lipowsky and Th.M.Nieuwenhuizen in
{\it Phase transitions and \hfill\break Critical Phenomena}, eds. C.Domb
and J.L.Lebowitz vol 14 (Academic Press, 1991) p.136.
\item{66.} X.Qiu, J.Ruiz-Garcia, K.J.Stine, C.M.Knobler and
J.V.Selinger, Phys.Rev.Lett {\bf 67}, 703 (1991);
S.Rivi\`ere, S. H\`enon and J.Meunier,Phys.Rev. E {\bf 49},
1375 (1994).
\item{67.} W.Helfrich, J.Phys. (Paris) {\bf 39}, 1199
(1978).
\item{68.} D.Nelson and J.Toner, Phys.Rev. B {\bf 24}, 363
(1981); J.Toner, Phys.Rev. B {\bf 26}, 462 (1982).
\item{69.} R.P.Rand, Ann.Rev.Biophys.Bioeng. {\bf 10}, 277
(1981).
\item{70.} R.Lipowsky and S.Leibler, Phys.Rev.Lett. {\bf
56}, 2541 (1986); R.Lipowsky, Europhys.Lett {\bf 7}, 255
(1988).
\item{71.} R.Ho\l yst and T.A.Vilgis, Europhys. Lett.{\bf 28}, 647
(1994).
\item{72.} M.Kl\'eman, J.Phys. (Paris) {\bf 35}, 595 (1974).
\item{73.} J.M.Kosterlitz and D.J.Thouless, J.Phys.C {\bf 6}, 1181 (1973);
J.M. Kosterlitz {\it ibid}, {\bf 7}, 1046 (1974).
\item{74.} R.Ho\l yst, Phys.Rev. B {\bf 50}, 12398 (1994).
\item{75.} P.G.de Gennes, Mol.Cryst.Liq.Cryst. {\bf 21}, 49 (1973).
\item{76.} J.D.Litster et al, J.Phys Colloq. (Paris) {\bf 40}, C3-339 (1979);
C.C.Huang,\hfill\break R.S.Pindak and J.T.Ho, Solid State Comm. {\bf 25}, 1015
(1978).
\item{77.} see e.g. A.Poniewierski and T.J. Sluckin, Liq.Cryst.
{\bf 2}, 281 (1987).
\item{78.} A.Allain and M.Kl\'eman, J.Phys. (France), {\bf 48}, 1799 (1987).
\item{79.} M.Kl\'eman
and C.E. Williams, J.Phys.Lett. (Paris) {\bf 35}, L-49 (1974).
\item{80.} P.G. de Gennes, Physics of Fluids {\bf 17}, 1645 (1974).
\item{81.} N.A.Clark, Phys.Rev.Lett. {\bf 40}, 1663 (1978).
\item{82.} E.Dubois - Violette, E Guazzelli and J.Prost, Phil.Mag.
{\bf 48}, 727 (1983).
\item{83.} L.Lejcek, Czech.J.Phys. B{\bf 32}, 767 (1982).
\item{84.} P.Oswald and L.Lejcek, J.Physique II {\bf 1}, 1067 (1991).
\item{85.} R.Bartolino and G.Durand, Phys.Rev.Lett. {\bf 39},  1346 (1977);
J.Phys.Lett. (Paris) {\bf 44}, L79 (1983).
\item{86.} R.Ribotta, Th\'ese d'Etat, Orsay (1975).
\item{87.} P.Oswald and D le Fur, C.R.Acad.Sci. Paris {\bf 297}, S\'erie II,
699 (1983).
\item{88.} P.Oswald and M.Kl\'eman, J.Phys.Lett (France) {\bf 45}, L319 (1984).
\item{89.} P.Oswald and M.Allain, J.Physique {\bf 46}, 831 (1985).
\item{90.} M. Allain, Europhys.Lett {\bf 2}, 597 (1986).
\item{91.} P.Oswald, C.R.Acad.Sci, {\bf 304}, 1043 (1987).
\item{92.} P.Oswald and M.Kl\'eman, J.Physique Lett. {\bf 43},L411 (1982).
\item{93.} P.Oswald, J.Physique Lett. {\bf 44}, L303 (1983).
\item{94.} P.Oswald, J.Physique {\bf 47}, 1091 (1986).

\centerline{\bf Figure captions}

\item{Fig.1} Schematic picture of smectic (lamellar) liquid crystal
obtained in computer simulation of a hard rod system by D.Frenkel.
\item{Fig.2} Lamellar phase in binary mixture of water and surfactant. The
surfactant molecuels form bilayers. From F.B.
Rosevear J.Soc.Cosmetic Chemist {\bf 19},
581 (1968).
\item{Fig.3} Schematic picture of lamellar phase in diblock copolymer system.
The linear copolymer consists of two homopolymers
A (solid line) and B (dashed line) joined by chemical bond (dot).
Alternating A rich and B rich domains are shown. After Ref.6.
\item{Fig.4} (a) The edge dislocation with small core, (b) The edge
dislocation with extended core (c)
The edge dislocation with core splitted into two disclinations
(d) The screw dislocation, after Ref.16
\item{Fig.5} (a) Schematic cross section of a sample containing a dislocation
array.
(b) strain versus position (c) accompanying variations of tilt angle versus
position
(d) Lagerwall subgrain boundary at the smectic A smectic C
phase transition. after Ref.13.
\item{Fig.6} Scan of the well annealed sample
in the lamellar phase showing regular
steps due to the edge dislocations. Each step corresponds to the termination of
the
single bilayer. The shot noise level is approximately 10\% of the step size.
after Ref.26.
\item{Fig.7} (a) AFM image of the holes and islands at the free surface of
AB diblock copolymer (here PS/PBMA diblock copolymer). The height of the
circular elevation is 310 \AA. (b) Cross section of the island. (c)
The schematic representation of the structure with the dislocation. Since
the islands are circular here we have dislocation loops. Locally the
distortions induced by large edge
dislocation loop are the same as those induced
by linear edge dislocation(Fig.4a).
The dark and white regions corresponds (see also Fig.3) to PS and PBMA domains.
The dark lines represent PS/PBMA interfaces, while dotted lines are fictive
PS-PS or PBMA-PBMA separations.
(d) TEM picture showing the edge dislocation. after Ref.21.
\item{Fig.8(a)} The idealized smectic meniscus of the freely suspended
smectic liquid crystal film, after Ref.23.
\item{Fig.8(b)} The arch-texture, after Ref.23.
\item{Fig.8(c)} Knots after Ref.23.
\item{Fig.9} Profiles of 8OCB droplets obtained by Michelson
interferometry at different temperatures. The indicated temperatures
are the numbers in of {}$^\circ$C below $T_{NA}$.
The apparent droplet diameter is $177\pm 5\mu$m . The matching
ofthe facet with the curved part of the droplet is tangential.
\item{Fig.10} Possible droplet configurations.
In (a) the  layers are parallel to the substrate. The top
surface shows a single facet parallel to the substrate and steps.
In (b) the top layer is curved to follow the free surface and dislocation
loops run through the bulk of the smectic liquid crystal.
\item{Fig.11} (a) Part of the ``goutte \`a gradins''
seen through themicroscope in reflection with natural light.
(b) Schematic view of a ``goutte \`a gradin'' (c)  Probable structure
of a ``gradin''.
\item{Fig.12} Droplet profiles of 4O.8 for different  temperatures
below $T_{NA}$ (droplet diameter 145$\mu$m). When the temperature
decreases the droplet shows a clear tendency to develop a discontinuity
in slope at the facet edge. A secondary facet  is well visible on the
profile at $T_{NA}-T=9.8^\circ$C.
\item{Fig.13} TEM picture of screw dislocations in lecithin
(courtesy of M.Allain). Arrows show the emergence points of screw
dislocations. Sample has been frozen and then fractured under vacuum.
A replica of its surface is observed  by TEM.
\item{Fig.14} The smectic order parameter $\epsilon$,
nematic order parameter $s$ and the local angle $\vartheta$, normalized by
their  bulk values, as a function of the distance from the center of the screw
dislocation core. The distance is measured in the units of the
correlation length
for smectic $\xi_\bot$. This length is of the size of a molecule
for small temperatures but diverges at the approach to the smectic-nematic
phase transition temperature
(in the case of continuous transition in liquid crystals$^{7}$).
After Ref.42.
\item{Fig.15} The screw dislocation tilted with respect to the layers.
\item{Fig.16} Two configurations of a dislocation pinned at the
solid substrate (a) unfavorable (b) favorable.
\item{Fig.17} The energy of a single edge dislocation
$F^*=(F(h)-E_0) 8\pi\lambda_s/
(\sqrt{KB} d^2)$
($\lambda_s =\sqrt{K_s/\sqrt{KB}}$)
versus the distance from the surface $h^*=2 h K/K_s$
for $\gamma/\sqrt{KB}=0$ (dashed line) and $\gamma/\sqrt{KB} =0.5$
(solid line). In the latter case the shallow minimum (not visible on the
scale of the Figure) is located at
$h^*=2.55$.
\item{Fig.18} The equilibrium location of the dislocation
inside the film $h_{eq}^{*}=h_{eq}/d$ versus the surface bending
elastic constant $K_s^{*}=K_s/Kd$ for $D/d=3$ -- long dashed line;
$D/d=10$ -- short dashed line; and $D/d=\infty$ -- solid line.
$\alpha_s=0$.
\item{Fig.19.} Clustering of dislocations in a wedge shaped sample
(Nallet and Prost$^{55}$).
\item{Fig.20} The edge profile $u^*=-u_{eq}(x,z=0)/d$,
versus
$x^*=(x-l)/\sqrt{\lambda h}$ at the free surface of thin lamellar film
deposited on solid substrate (dashed line). Here $K_s=K d$,
$\gamma/\sqrt{KB}=5$, $D=3d$, $h=d/2$. For comparison (solid line)
we show the profile, obtained in the previous
approach$^{19}$ with finite size and surface effects
neglected ($D\to\infty$, $\gamma =0$, $K_s=0$).
\item{Fig.21} A plot of the local film thickness against position measured
along an axis parallel to the substrate and normal to the domain edge
(see Fig.5) (a) the average film thickness, $D$ is $d<D<2d$;
(b) $2d<D<3d$; (c) $3d<D<4d$; (d) $5d<D<6d$. Dots display experimental results
and the solid line was obtained from the theory (however here $K_s=0$).
After Ref.22.
\item{Fig.22} The same schematic picture as shown in Fig.7 with steps at the
free surface, without dislocations in the bulk. (this configuration is in fact
unfavorable in comparison to film structure with steps induced by
dislocations Fig.5 see Section VI). After Ref.60.
\item{Fig.23} The edge profile $u^*=-u(x,z=0)/d$ versus
$x^*=(x-l)/\sqrt{\lambda d}$ for the film on a solid substrate bounded by the
interface
with the surface tension $\gamma=0$ for
three thicknesses of the lamellar film: $D/d=3$ -- long dashed line;
$D/d=10$ -- short dashed line; and $D/d=\infty$ -- solid line.
(a) $K_s=K d$ (b) $K_s/Kd=10$.
\item{Fig.24} The top view of the $z=0$ surface
containing the edge dislocation. The dashed line represents
the unperturbed position of the dislocation and $u_x(y)$ is
the amplitude of thermal distortions of the dislocation in
the x-y plane (see also Eq(9.1)).
\item{Fig.25} The dimensionless interaction energy
per unit length $V(x)=2\pi^2\sqrt{N}/(\sqrt{KB}d)
\hfill\break (F_{s}(x)+F_H(x)-F_{s}(0))$ versus the separation
$x=l/\sqrt{\lambda dN}$ for $N=30$,$N=40$ and $N=54$.
Here $N$ is the number of lamellar layers. As $N$ increases
the minimum of $V(x)$ (at $x_0\approx 3$) moves upward.
At $N=54$ $V(x_0)=V(x=\infty)=0$ and the unbinding
transition occurs ($K_s/Kd=1$, $\alpha_s=1$).
\item{Fig.26} The surface bending elastic constant
$K_s^*=K_s/Kd$ ($\alpha_s=1$) versus the number of
lamellar layers, $N$, at the unbinding transition.
The
dotted line divides the diagram into two regions.
Above the line the dislocations are
free and below they are bound.
\item{Fig.27} The surface tension $\alpha_s=\gamma
/\sqrt{KB}$ ($K_s/Kd=1$) versus the number of lamellar layers, $N$
at the unbinding transition. The dotted line
divides the diagram into two regions. Above the line
the dislocations are free, below they are bound.
At the line the unbinding transition occurs.
\item{Fig.28} The climb motion of an edge dislocation
submitted to a normal stress $\sigma_{zz}^E$.
$v$ is the velocity, ${\bf b}$ is the Burgers vector and ${\bf t}$ is the
unit vector along the dislocation.
\item{Fig.29}  (a) (b) climb of the giant dislocation and equivalent
rheological model.
\item{Fig.30} The glide motion of a dislocation
submitted to shear
(see Fig.29).
\item{Fig.31} Thermally activated bulge.
\item{Fig.32} The schematic view of the dynamical measurements.
(a) step like deformations. Initial thickness variation is $\delta_0$.
(b) Normal stress measured  experimentally.
The characteristic relaxation time is $\tau$.
(c) Time $\tau$ versus $\sigma_0$ in a 400$\mu$m thick sample of 8CB
at room temperature.
\item{Fig.33} Helical instability of screw dislocation.
\item{Fig.34} Crossing of an edge and a
screw dislocation with the
formation of the jog.
\item{Fig.35} Lubrication geometry.
\item{Fig.36} The apparent (effective) viscosity versus the shear rate
(8CB, $D=100\mu$m, 22$^\circ$C from Ref.91).
\item{Fig.37} Velocity profile in the sample and a dislocation
distorted by the flow.
\vfill\eject\end